\pdfoutput=1
\documentclass[11pt]{article}

\usepackage[final]{acl}

\usepackage{times}
\usepackage{latexsym}
\usepackage[T1]{fontenc}
\usepackage[utf8]{inputenc}
\usepackage{microtype}
\usepackage{inconsolata}
\usepackage{graphicx}
\usepackage{booktabs}
\usepackage{multirow}
\usepackage{array}
\usepackage{amsmath,amssymb}
\usepackage{xcolor}
\usepackage{tikz}
\usetikzlibrary{positioning}
\usepackage{url}

\definecolor{cardA}{HTML}{E8F0FE}
\definecolor{cardB}{HTML}{FFF3E0}
\definecolor{cardborderA}{HTML}{4A6FA5}
\definecolor{cardborderB}{HTML}{C77E2C}

\title{PriceBench: A Diagnostic Benchmark for Price, Quality, and Brand Preferences in LLM Booking Agents}

\author{Pavel Kireyev \\
  Department of Management \\
  London School of Economics and Political Science \\
  \texttt{p.kireyev@lse.ac.uk}}

\begin{document}
\maketitle

\begin{abstract}
LLMs increasingly act as purchasing agents, which makes the LLM, not the user, the one choosing among the options that satisfy a request; its preferences quietly fix what gets bought and what it costs. Hotel booking is a clean instance: a high-volume choice settled on a few comparable attributes, where the pick reveals those preferences. We introduce PriceBench, a diagnostic benchmark that recovers an LLM's price, quality, and brand preferences from its booking choices with a logit choice model, applied to 28 LLMs from 8 providers on 3{,}600 hotel tasks from 179 real New York City properties. We find that capability is associated with how consistently an LLM chooses, not with what it chooses: more capable LLMs hold stronger, more consistent preferences, while weaker ones either lock onto one position, exploitable by whoever controls listing order, or choose almost indifferently. What those preferences favor varies sharply across providers and even within one family: price sensitivity spans more than an order of magnitude, and the price/quality trade-off moves mean booked nightly price from \$247 to \$393 on identical tasks. What an agent buys must therefore be measured per LLM, not inferred, and we release the tasks, code, and all 28 response sets.
\end{abstract}

\section{Introduction}
\label{sec:intro}

Large language models are increasingly deployed as autonomous agents in user-facing commerce: shopping assistants, travel concierges, and customer-support bots that complete transactions on behalf of users \citep{bansal2025magentic,wang2025ecombench,xie2024travelplanner,yao2022webshop,zhou2024webarena}. When such an agent receives a query with explicit guardrails (a budget cap, a minimum star rating, a preferred neighborhood), the request typically admits several listings that meet the constraints. What tips the agent one way or the other is its \emph{implicit} taste for price, quality, brand, and a dozen things the user never specified. Existing agent benchmarks measure whether the agent completes such a task; they do not measure \emph{which} of the admissible options the agent picks, and therefore do not measure the agent's revealed purchasing policy \citep{filandrianos2025biasbeware,fish2025econevals,kamruzzaman2024brandbias,ross2024llmeconomicus}.

\begin{figure*}[t]
\centering
\newcommand{\hotelcard}[7]{%
\scalebox{0.7}{%
\begin{tikzpicture}[
    headerA/.style={fill=cardborderA, draw=cardborderA,
                     rounded corners=3pt, inner sep=0pt},
    headerB/.style={fill=cardborderB, draw=cardborderB,
                     rounded corners=3pt, inner sep=0pt},
    bodyA/.style={fill=cardA, draw=cardborderA, line width=0.5pt,
                   rounded corners=3pt, inner sep=8pt},
    bodyB/.style={fill=cardB, draw=cardborderB, line width=0.5pt,
                   rounded corners=3pt, inner sep=8pt}
]
\node[body#1, align=left, text width=7.2cm, anchor=north] (card)
{%
    \vspace{2pt}%
    {\color{cardborder#1}\sffamily\fontsize{7.5}{9}\selectfont\bfseries OPTION #1}\\[2pt]
    {\sffamily\fontsize{10}{11}\selectfont\bfseries #2}\\[5pt]
    {\fontsize{8.5}{11}\selectfont
    \begin{tabular}{@{}p{2.7cm}@{}p{3.9cm}@{}}
    Star rating       & #3 \\
    Neighborhood      & #4 \\
    Guest reviews     & #5 \\
    Room type         & Standard King \\
    Free cancellation & Yes \\
    Breakfast         & No \\
    Amenities         & #6 \\
    \end{tabular}}\\[3pt]
    {\color{cardborder#1!50}\rule{6.8cm}{0.4pt}}\\[2pt]
    {\sffamily\fontsize{10}{11}\selectfont\bfseries Price per night: \$#7}%
};
\end{tikzpicture}}%
}
\hotelcard{A}{voco Times Square South}{3\,$\star$}{Hell's Kitchen}%
{8.0/10 (1{,}406)}{WiFi; Restaurant; Fitness}{148}%
\hspace{6pt}%
\hotelcard{B}{Kimpton Theta New York}{4\,$\star$}{Times Square}%
{8.2/10 (697)}{WiFi; Wine hour; Fitness}{281}
\caption{Task 113 from PriceBench: \emph{which option do you choose?} GPT-5.4, GPT-5.4 Mini, GPT-4.1 Mini, Claude Haiku 4.5, and Gemma3~27B take A; Gemma3~4B, DeepSeek-R1 7B, and Phi-4 Mini take B. Different agents, different revealed preferences.}
\label{fig:example}
\end{figure*}

We introduce PriceBench, a diagnostic benchmark in the spirit of behavioral litmus tests for LLM economic decision-making \citep{fish2025econevals,ross2024llmeconomicus}. The benchmark presents 3{,}600 hotel choice tasks (1{,}800 binary, 1{,}800 ternary) drawn from 179 real NYC properties spanning 1--5 stars, 36 neighborhoods, and \$45--\$1{,}650 per night. Each task is scored twice per LLM, once in the original ordering and once with the options swapped. We fit a logit choice model \citep{mcfadden1974conditional,train2009discrete}: the chosen option is regressed on the difference in each attribute between the options, with an intercept that absorbs any constant preference for the first-listed slot. Each preference parameter is then identified from how the LLM's choices move with attribute differences, both between the two options in a task and across the re-randomized repetitions of each pair. Figure~\ref{fig:example} shows one such task and how the LLMs split on it.

Applying PriceBench to 28 LLMs from 8 providers, we find that capability is associated with the consistency of an LLM's choices but not with their content: more capable LLMs hold stronger, more consistent preferences, their bookings reliably driven by price and quality, while weaker ones fix on one position regardless of the options or choose close to indifferently. Five of 28 LLMs are \emph{position-locked} at temperature zero, picking the first option regardless of attributes, and sampling recovers no usable preferences from them (\S\ref{sec:robustness}), so any platform that controls listing order controls 88--100\% of their bookings (\S\ref{sec:results_triage}). Among the 23 that do engage, how sharply an LLM weighs attributes is partly associated with scale, but which way it leans, toward price or toward quality, is not: willingness to pay for a one-point gain in review score spans 16$\times$ and splits same-family ladders, moving mean booked nightly price from \$247 to \$393 on identical tasks (\S\ref{sec:results_price}--\S\ref{sec:results_personality}). Brand preferences vary too: they persist after controlling for price and quality, yet are unrelated to the provider (permutation $p = 0.48$) (\S\ref{sec:results_brand}). On the LLM subsets retested, the price and quality readings survive prompt reformatting, sampled decoding, and five-option lists, though the dollar magnitudes are format-conditional (\S\ref{sec:robustness}), and the engaged LLMs value quality above the range documented for human bookers (\S\ref{sec:results_human}). More broadly, these preferences are a surface the sell side can act on: a platform can tune rankings and prices to an engaged agent's utility, or simply reorder listings to exploit a position-locked one.

What an agent actually buys is therefore specific to the individual LLM and has to be measured, not assumed; PriceBench provides that reading directly from the LLM's booking decisions. We release the 3{,}600 tasks, the estimation code, all 28 response sets, and the prompt-variant, decoding, and five-option runners built for \S\ref{sec:robustness}, so that any new LLM can be audited against the same set in a single scoring pass.\footnote{\url{https://github.com/Pashasan/pricebench-emnlp}}

\section{Related Work}
\label{sec:related}

\paragraph{Agentic commerce benchmarks score task success; we measure the choice.}
WebShop \citep{yao2022webshop}, WebArena \citep{zhou2024webarena}, and TravelPlanner \citep{xie2024travelplanner} test whether an agent can navigate a site or satisfy budget constraints, and ECom-Bench \citep{wang2025ecombench} and $\tau$-bench \citep{yao2024taubench} score customer-support success at low pass rates (10--20\% and under 50\%, respectively). Magentic Marketplace \citep{bansal2025magentic} goes further, simulating a two-sided market and documenting a severe first-proposal bias that hands a 10--30$\times$ advantage to response speed over quality. None of them isolates which qualifying option an agent actually prefers, the step at which its purchasing policy is revealed; that is what PriceBench measures. Ordering sensitivity is itself well documented \citep{liu2024lost,pezeshkpour2024order,zheng2023judging,zheng2024robust}; we fold it in as a prerequisite, treating Magentic's first-proposal effect as a gate an LLM must clear before its preferences are identifiable at all (\S\ref{sec:results_triage}).

\paragraph{Recovering price preferences from choices.}
Price sensitivity and willingness to pay are the classic targets of discrete-choice and conjoint estimation \citep{mcfadden1974conditional,train2009discrete}, in which price enters as an attribute and dollar valuations fall out as ratios of coefficients. A growing line applies this lens to LLMs as economic subjects: \citet{horton2023homosilicus} treat them as simulated decision-makers, \citet{ross2024llmeconomicus} and \citet{fish2025econevals} map their biases and separate capability from behavioral disposition (a split that parallels ours), \citet{goli2024preferences} recover discount rates from LLM choices by maximum likelihood, and \citet{wang2025augmentation} use LLM conjoint responses as synthetic market research. All treat the LLM as a stand-in for a human respondent, which \citet{gui2025causal} caution against. We instead make the LLM's own purchasing policy the object: from its bookings we recover price sensitivity, its functional form, and dollar willingness to pay across 28 LLMs, and read brand preference, shown before only in text \citep{kamruzzaman2024brandbias}, off the same binding choices, asking throughout whether any of it is predictable from the LLM's provider (\S\ref{sec:results_brand}). These defaults are what an agent runs on once it is the buyer and, as \citet{fish2024collusion} show for interacting pricing agents, what aggregate into market outcomes.

\section{Benchmark Design}
\label{sec:design}

\paragraph{Hotel pool.}
Hotel booking is a high-volume, constrained-choice task that agentic systems increasingly target \citep{xie2024travelplanner}: business and corporate travel especially is booked repeatedly against a small, standardized set of comparable attributes (price, location, star rating, reviews). That makes a single booking clean to model while remaining a realistic agent task. It is also where we expect agentic commerce to arrive first: delegation is already the norm through corporate travel desks and online travel agencies (OTAs), the attributes are standardized and machine-legible, meeting a user's constraints is easy to verify, and the sell side already prices by algorithm. We constructed a frozen pool of 179 real NYC hotel profiles from listings on four OTAs (Booking.com, Expedia, KAYAK, TripAdvisor), spanning 1--5 stars, 36 neighborhoods, and \$45--\$1{,}650 per night. Each hotel carries ten attributes: four \emph{fixed} (star rating, neighborhood, chain affiliation, amenities), constant across appearances, and six \emph{re-randomized} on each appearance. The re-randomized draws are a fresh price (uniform within the hotel's listed range), room type, free-cancellation and breakfast flags, and small perturbations of the hotel's listed (\emph{base}) review score (within $\pm 0.2$) and review count (within $\pm 10\%$). Re-randomization also makes the repeated presentations of a pair non-identical, so each attribute varies both between the two hotels and across repetitions, which is what lets the difference-based estimator (\S\ref{sec:method}) recover its weight.

\paragraph{Tasks.}
We sample 450 unique hotel pairs (each presented 4 times, $=1{,}800$ binary tasks) and 300 unique triples (each presented 6 times, $=1{,}800$ ternary tasks). The binary pairs carry the main analysis; the ternary triples are a robustness check that the recovered preferences are not an artifact of the two-option format (Appendix~\ref{app:price_bt}). Every task is scored twice (once in the original ordering and once with options swapped), yielding up to 7{,}200 observations per LLM. Pure position bias cancels in the pooled estimate.

\paragraph{LLMs.}
We score 28 LLMs from 8 providers (OpenAI, Anthropic, Google, Meta, Alibaba, Microsoft, Mistral, DeepSeek), spanning sub-billion-parameter open-weight LLMs to frontier proprietary LLMs. All runs use temperature 0 and greedy decoding (API: \texttt{temperature=0, top\_p=1}). Prompts present options as Booking.com-style cards with all ten attributes labeled (Appendix~\ref{app:prompt}).

\section{Estimation}
\label{sec:method}

\paragraph{Choice model.} For each LLM we fit a logit choice model to its bookings. On the binary tasks, which carry the main analysis, this is a logistic regression of the chosen option on the difference in each attribute between the two options,
\begin{equation}
\label{eq:utility}
\begin{split}
\Pr(A) = \sigma\big(c &+ f(p_A)-f(p_B) \\
&+ (\mathbf{x}_A-\mathbf{x}_B)^{\top}\boldsymbol{\gamma}\big),
\end{split}
\end{equation}
where $\sigma$ is the logistic function, $\mathbf{x}$ collects the non-price attributes, $f(p)$ is the price term, and the intercept $c$ absorbs any constant tendency to pick the first-listed option (the position bias of \S\ref{sec:results_triage}). Because the regressors are attribute differences and every task is also scored with the options swapped, a pure position effect loads on $c$ and leaves the attribute weights $\boldsymbol{\gamma}$ uncontaminated. This is the binary case of a discrete choice model \citep{mcfadden1974conditional,train2009discrete}; for the three-option tasks we fit the corresponding multinomial logit over the three alternatives (Appendix~\ref{app:price_bt}).

\paragraph{Why the price functional form matters.} We estimate three forms for $f(p)$, and which one an LLM follows is itself deployment-relevant: it sets where on the price scale the agent is sensitive, and so how a seller can move it. \emph{Linear}, $f(p)=\beta_{\mathrm{lin}}\,p$, is a constant disutility per dollar; \emph{log}, $f(p)=\beta_{\ln p}\ln p$ (Weber--Fechner perception), makes that sensitivity diminish as price rises; \emph{non-parametric}, $f(p)=\sum_{k=2}^{10}\delta_k D_k(p)$ with $D_k$ a price-decile indicator, leaves the shape free and exposes thresholds and price-as-quality effects a ranking or pricing algorithm could exploit. Decile cut-points are computed once on the pooled price distribution and reused across LLMs for comparability.

\paragraph{Brand specification.} For the brand audit (\S\ref{sec:results_brand}) we add chain-family dummies (Hilton, Marriott, IHG, Hyatt, Wyndham; Independent is the reference) to the price-decile, star, room-type, cancellation, breakfast, and review controls, so each chain's coefficient is its choice premium over an otherwise-comparable independent hotel (Appendix~\ref{app:brand}, Figure~\ref{fig:brand}).

\section{Results}
\label{sec:results}

The results build in one direction: capability is associated with how \emph{decisively} an LLM chooses, not with \emph{what} it chooses. An LLM must first engage with the options at all (\S\ref{sec:results_triage}); among those that do, we turn first to price, and to how strongly they weigh it (\S\ref{sec:results_price}). The direction of the policy, how an LLM trades price against quality (\S\ref{sec:results_personality}) and how it treats brand (\S\ref{sec:results_brand}), shows wide heterogeneity, associated with neither scale nor provider, that must be measured for each LLM; \S\ref{sec:results_human} and \S\ref{sec:robustness} then benchmark the recovered valuations against the human record and test their stability.

\subsection{Position engagement is the gateway test}
\label{sec:results_triage}

A purchasing policy presupposes that the LLM reads the options at all, so we screen on that first. The first-shown rate scores, over all 3{,}600 tasks in both orderings, how often an LLM picks whichever of a task's two reference options was listed earlier (on binary tasks, the first-listed option; exact construction in Appendix~\ref{app:triage}): 50\% is fully attribute-driven, 100\% is pure slot-following. We flag any LLM outside an \emph{ex ante} band of $[15\%, 85\%]$ as position-locked; the verdict is cutoff-insensitive, since no LLM falls in the 80.7--88.2\% gap between the highest engaged and the lowest locked rate (Appendix~\ref{app:triage}).

Five LLMs lock at temperature zero: Llama 3.2 1B (100.0\%), Mistral 7B (99.5\%), Qwen3 0.6B (98.5\%), and Llama 3 and 3.1 8B (both 88.2\%). The lock is a failure mode that disqualifies preference estimation, not a preference we attribute to the LLM: every locked LLM answers validly and follows the shown slot under both orderings, so it is not a parsing failure (Appendix~\ref{app:triage}), and sampling either preserves it or dissolves it into attribute-poor noise (\S\ref{sec:robustness}). Whatever its cause, these LLMs behave as constants and are directly exploitable: any platform that controls listing order controls 88--100\% of their bookings. The lock skews small and weak but is not a clean size threshold: two 8B LLMs lock while smaller ones (Llama 3.2 3B, Gemma3~4B) engage. The remaining 23 clear it and carry interpretable preferences; their milder within-band ordering leanings are cataloged in Appendix~\ref{app:triage}.

\subsection{Price sensitivity and functional form}
\label{sec:results_price}

\begin{figure*}[!tb]
\centering
\includegraphics[width=0.88\textwidth]{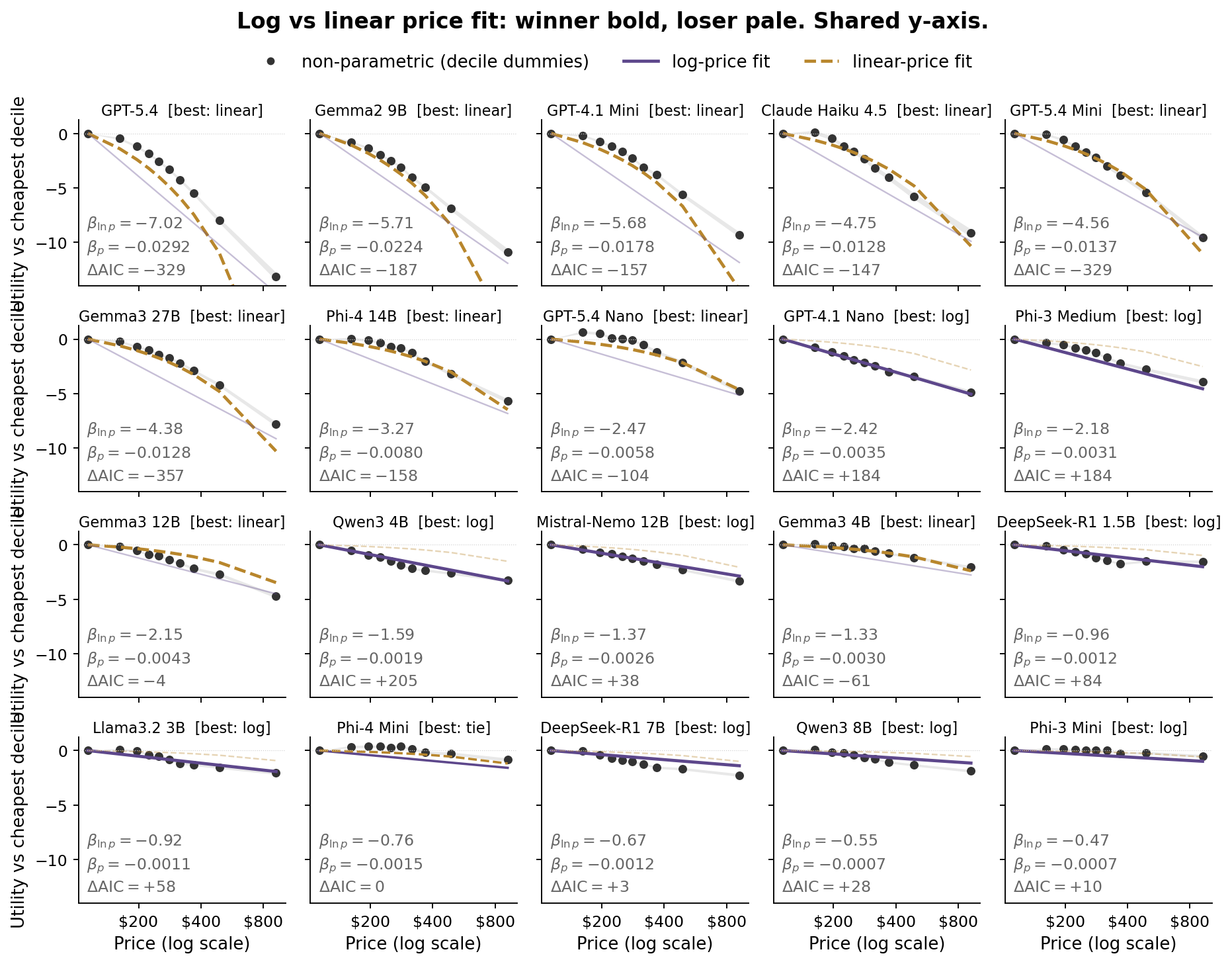}
\caption{Per-LLM price response. The $y$-axis is each price decile's choice utility relative to the cheapest decile, in logit (log-odds) units: $0$ = as likely to be chosen as the cheapest rooms, more negative = less likely; all three estimators are pinned to $0$ there. Black points are non-parametric decile estimates ($\pm 1$ SE band); the log-price (purple) and linear-price (orange dashed) curves are fit to the raw choices, not to the 10 decile points, so they need not pass through them, and the AIC winner is drawn bold. Each panel reports, bottom-left, $\beta_{\ln p}$, $\beta_p$, and $\Delta\mathrm{AIC} = \mathrm{AIC}_{\text{lin}} - \mathrm{AIC}_{\text{log}}$ (negative = linear wins); panels are sorted by $|\beta_{\ln p}|$, most sensitive first. Twenty of the 21 engaged LLMs with a genuine negative price response are shown; the weakest, Qwen3~30B-A3B ($\beta_{\ln p}=-0.40$, AIC favors log), is omitted for space but appears in the full 23-panel grid (Appendix~\ref{app:price_decile}), with the two flat-response LLMs.}
\label{fig:funform}
\end{figure*}

Price is the attribute a seller most directly controls. An agent's price response varies widely across LLMs in both strength and shape, two deployment-relevant properties. Log-price coefficients span more than an order of magnitude across the engaged set: $\beta_{\ln p} = -7.02$ for GPT-5.4 at one extreme and $-0.40$ for Qwen3~30B-A3B at the other. Two LLMs do not respond to price in any deployment-relevant sense (Gemma3 1B, Phi-2 2.7B); every other engaged LLM has a negative coefficient significant at $p<0.001$.

Figure~\ref{fig:funform} overlays the per-LLM parametric fits on the non-parametric decile points, each panel comparing the three nested estimators on the same axes. The non-parametric shapes vary substantially: sharp linear-like declines (GPT-5.4, Gemma3~27B), convex shapes (Claude Haiku 4.5), threshold patterns (indifferent up to $\sim$\$200, then dropping; Llama 3.2 3B, Gemma3~4B), and an \emph{early positive bump} in several LLMs where utility briefly rises above the cheapest decile, consistent with price-as-quality-signal inference; GPT-5.4 Nano shows the clearest such bump ($+0.62$ at decile 2). Each shape is a distinct lever: threshold agents can be nudged across their cliff, bump agents reward a seller that avoids being cheapest, linear agents yield only to absolute discounts.

Of the 21 LLMs with a genuine negative price response, AIC selects linear for 10 and log for 10 (one tie), and the decisively linear and most-price-sensitive rankings agree at the top: the same eight LLMs occupy the top of both lists (GPT-5.4, GPT-5.4 Mini, Gemma3~27B, Gemma2~9B, Phi-4 14B, GPT-4.1 Mini, Claude Haiku 4.5, GPT-5.4 Nano). The two forms split the agents by where a discount bites: a linear response (the most price-sensitive LLMs) gives a fixed-dollar cut the same pull at \$100 or \$800, while a log response (the more moderate ones) rewards only proportional cuts, biting hardest at the low end of the price range.

\subsection{The price--quality map}
\label{sec:results_personality}

\begin{figure*}[!tb]
\centering
\includegraphics[width=0.66\textwidth]{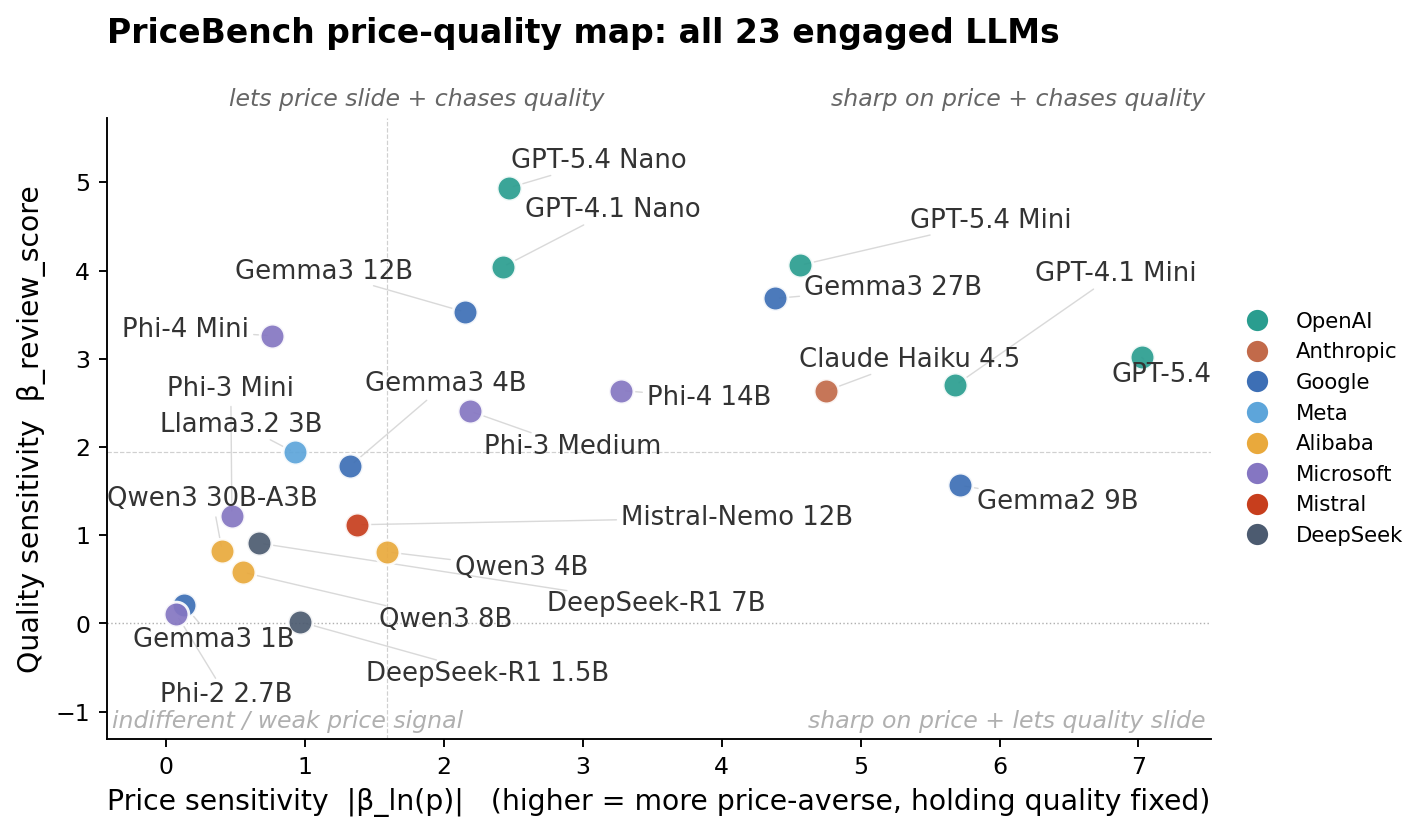}
\caption{Price--quality map. $x$-axis: $|\beta_{\ln p}|$ (price sensitivity, holding quality fixed). $y$-axis: $\beta_{\text{review\_score}}$ (quality sensitivity, holding price fixed). Colored by provider; dashed lines at the median split the plane into four regions. Distance from the origin is largely \emph{decisiveness} (the logit scale inflates every coefficient at once), which is why most LLMs fall on the diagonal; the scale-free price/quality trade-off is shown in Figure~\ref{fig:wtp} (Appendix~\ref{app:quality_full}).}
\label{fig:personality}
\end{figure*}

Clearing the gate and weighing price still leaves the central trade-off: when several rooms fit the request, does an LLM take the cheaper one or the better one? Engaged LLMs split widely, and Figure~\ref{fig:personality} maps each by how strongly it weighs price ($x$-axis) against quality ($y$-axis), the latter summarized by the review-score coefficient, clearest of the five quality signals we estimate (full set in Appendix~\ref{app:quality_full}).

Read naively, the map mostly reflects \emph{decisiveness}: because the logit scale inflates every coefficient at once, an LLM that chooses consistently posts large weights on price and quality alike, while one whose temperature-zero choices are only loosely tied to attributes posts small weights on both, so LLMs line up along the diagonal ($r = 0.55$; ten read both signals sharply, ten neither). The same force drives the engagement gate (\S\ref{sec:results_triage}) and the price-strength scaling (Appendix~\ref{app:price_scaling}): more capable LLMs choose more decisively, and that decisiveness must be divided out before the price-versus-quality lean, what an LLM actually prefers, comes into view.

That lean is already on the map, distinct from distance: where distance from the origin is decisiveness, the lean is the point's \emph{slope}, the ratio of quality weight to price weight ($y/x$ in Figure~\ref{fig:personality}). Taking the ratio cancels the overall coefficient scale, stripping out decisiveness; LLMs on a steeper ray are quality-chasers, those on a shallower ray bargain-hunters. In dollars the same ratio is willingness to pay for one review-score point. Because a log-price response makes a dollar's value depend on the price level, we read it at a reference nightly rate of \$250, a typical listed price: $\mathrm{WTP} = (\beta_{\text{review\_score}}/|\beta_{\ln p}|)\times\$250$, ranked per LLM with standard errors in Appendix~\ref{app:quality_full} (Figure~\ref{fig:wtp}). The spread is wide: across the 20 LLMs with well-identified coefficients it runs roughly 16$\times$, from \$68 (Gemma2~9B) and \$107 (GPT-5.4) at the bargain-hunter end to \$1{,}071 (Phi-4 Mini) at the quality-chaser end (cluster-bootstrap 95\% CI on the ratio $[11.7\times, 23.7\times]$; Appendix~\ref{app:bootstrap}). It is uncorrelated with decisiveness ($r = -0.09$): how decisively an LLM chooses says nothing about which way it leans. It even splits siblings: down the GPT-5.4 line, willingness to pay climbs from \$107 to \$222 (Mini) to \$500 (Nano), so the smallest sibling values a quality point almost five times as highly as the largest. Stepping down a family ladder changes what the agent prefers, not just how sharply it chooses. The dollar levels themselves are conditional on the elicitation format, which moves them systematically while preserving the ordering (\S\ref{sec:robustness}).

These leanings show up directly in what gets booked. Across the 3{,}600 binary tasks, GPT-5.4 books at a mean of \$247 per night while Phi-4 Mini books at \$393, a \$146 spread on the same task set: GPT-5.4 anchors the cheap end (strong price weight), Phi-4 Mini the expensive end (weak price weight, heavy quality loading). Among the most decisive LLMs the spread is narrower but still material (\$247 to \$346, $\sim$\$99). What an identical set of listings sells for thus depends on which LLM is shopping it.

\subsection{Brand preferences across LLMs}
\label{sec:results_brand}

Brand is another attribute LLMs differ on. Holding price and the measured quality signals fixed, we estimate each LLM's preference for the five hotel chains relative to an independent hotel (Appendix~\ref{app:brand}, Figure~\ref{fig:brand}). These preferences are real and, like price and quality, vary sharply from one LLM to the next: 11 of the 23 engaged LLMs prefer all five chains to an otherwise-identical independent, and they disagree on which chains and by how much. Part of the pull is location: neighborhood controls cut the all-five-chains count from 11 to 7 of 23 and the largest premium, GPT-5.4's Wyndham coefficient, from $+2.17$ to $+1.23$, still significant and spread across all five Wyndham properties rather than resting on one (Appendix~\ref{app:brand}). Whether the residual pull is preference or brand association absorbed in pretraining is not something behavior alone can settle; for deployment the consequence is the same, since either way the booking lands on the chain.

This heterogeneity holds within a provider as much as across it: two LLMs from the same provider are, on average, no more alike in brand preference than two from different providers (mean pairwise correlation of brand-coefficient vectors $+0.079$ within versus $+0.090$ across; one-sided $p = 0.48$ over 2{,}000 permutations of the provider labels), and even the GPT-5.4 siblings, alike on nearly everything else, diverge here. The null survives every specification we tried ($p$ between 0.42 and 0.63; Appendix~\ref{app:brand}).

\subsection{LLM valuations against the human record}
\label{sec:results_human}

The dollar figures above need a reference point, and the hospitality and choice-modeling literatures provide one, with the caveat that the human benchmark is a distribution, not a number: price sensitivities vary severalfold across households and tourist segments \citep{rossi1996value,berry1995automobile,masiero2012segmentation}. For one point on a ten-point review scale at a \$250 nightly rate, studies of how review scores move posted hotel prices, a research-center report and a working paper, put humans at roughly \$2--\$22 \citep{anderson2012lodging,lewis2016welfare}; structural models of real bookings run from zero \citep{ursu2018rankings} to roughly \$30 per platform rating point and \$61 summed across platforms (our conversions of \citealp{ghose2012ranking}), with the entire premium for a rating above 4.5 over one below 4 estimated at \$99 per night \citep{chen2017sequential}; Appendix~\ref{app:anchor} documents every conversion. Against that range the 20 well-identified engaged LLMs are high: they value the same point at \$68 to \$1{,}071 (median \$269), so even the most frugal sits above the top of the documented per-point range, and the median several times above it. The constructs differ, and human choices on the PriceBench tasks themselves remain uncollected (Limitations); the comparison bounds plausibility rather than fixing a correct value. Still, a gap this size is exactly why an agent's valuations must be measured, not assumed human-like.

\section{Robustness}
\label{sec:robustness}

The estimates above come from one elicitation: a neutral prose card, a single-letter answer, two or three options, greedy decoding. We vary each in turn; the runners are in the release, and Appendices \ref{app:variants}--\ref{app:bootstrap} carry full tables.

\paragraph{Prompt format and reasoning.}
We crossed four variants (JSON record cards, attribute rows reordered price-first, a mild client persona, and chain-of-thought elicitation with free-text reasoning before the answer) with six LLMs spanning the willingness-to-pay range, GPT-5.4 Nano and Mini included, in both orderings at temperature zero (Appendix~\ref{app:variants}). No LLM's price--quality lean flips in any cell where both coefficients remain estimable: review coefficients stay positive in every cell, price coefficients negative in all but one (Phi-4 Mini under JSON, $+0.03$, SE $0.14$). The cross-model ordering holds in every variant, never below Spearman $+0.60$ against the baseline, and $+0.89$ or higher in the three single-letter variants. Dollar magnitudes do move with format, in two systematic directions: the persona raises the quality lean of all six LLMs (Gemma2~9B, \$70 to \$215), and chain of thought lowers it in all five that ran it (Phi-4 Mini, \$1{,}020 to \$147). Format moves how sharply an LLM chooses and the dollars that follow, not which way it leans: the scale-versus-taste split \citep{fiebig2010gmnl}.

\paragraph{Decoding.}
Sampling at temperature 0.7 does not undo the locks: four of the five still take the first-shown option 89.2--99.3\% of the time; the fifth, Qwen3 0.6B, loosens to 52.2\% but carries almost no attribute signal (pseudo-$R^2 = 0.031$, well below both engaged controls), noise around a positional habit. The engaged controls keep their signs (Appendix~\ref{app:temp}); greedy decoding is not manufacturing the lock, and no locked LLM yields usable preferences under sampling.

\paragraph{Larger choice sets.}
The ternary block already reproduces the binary price ranking (Spearman $\rho = 0.88$--$0.90$; Appendix~\ref{app:price_bt}), and on 300 new five-option tasks from the same pool, utilities fitted on the binary block predict the five-way pick out of sample (hit rates 42--52\% vs.\ 20\% chance, top-two 66--84\% vs.\ 40\%). While engaged LLMs spread their picks across slots (2--10\% first-slot), the boundary LLM Mistral-Nemo 12B (55.5\%) and the locked Llama 3.1 8B (61.2\%) carry the position habit into longer lists (Appendix~\ref{app:fiveopt}).

\paragraph{Statistical stability.}
A cluster bootstrap resampling the 450 binary choice pairs whole ($B = 500$, no failed refits) reproduces the reported orderings (mean rank correlation 0.996 and 0.991 for price sensitivity and willingness to pay), and every coefficient claimed at $p<0.001$ survives Bonferroni and Benjamini--Hochberg over all 184 per-LLM tests, the largest claimed raw $p$ being $7.7\times10^{-7}$ (Appendix~\ref{app:bootstrap}).

\paragraph{External capability axis.}
Parameter counts and pricing tiers are partial capability proxies, so we re-tested the capability result on a public index that scores open and proprietary LLMs on one scale \citep{artificialanalysis2026}. Across the 18 engaged LLMs it covers, both halves hold: the index correlates with how decisively an LLM chooses ($r = +0.63$ with decisiveness, $+0.71$ with price-coefficient magnitude; $p = 0.005$ and $0.001$) but not with which way it leans (willingness to pay, all $p > 0.09$). Appendix~\ref{app:price_scaling} details the caveats.

\section{Conclusion}
\label{sec:conclusion}

\looseness=-1
PriceBench recovers an LLM's price, quality, and brand preferences from its booking choices with a logit choice model. Across 28 LLMs from 8 providers, capability is associated with how \emph{decisively} an LLM chooses, not with what it chooses: 5 of 28 lock onto the first listing (88--100\%), and among the rest the price--quality balance is unrelated to capability, splits same-provider siblings, and shifts mean booked price by \$146; brand preferences vary just as much (permutation $p = 0.48$). What an LLM buys is thus a property of that LLM, not its size or provider: a seller can price and rank to exploit it; a deployer must measure it for each LLM. Auditing a new LLM is a single scoring pass: a routine pre-deployment check.

\section*{Limitations}

PriceBench measures preferences within a deliberately narrow slice of decisions; several extensions are natural follow-ups rather than rebuttals. \emph{Domain.} All tasks involve NYC hotel rooms, and the methodology is validated only there; price sensitivity likely varies across product categories (groceries, flights, enterprise software), and the difference-based estimator assumes attribute variation comparable to what hotels naturally provide. Porting the benchmark to another category needs only a new profile pool, but we claim no validation beyond hotels. \emph{Elicitation.} The headline estimates are conditional on a neutral booking prompt at temperature zero. \S\ref{sec:robustness} shows that the price--quality lean and the cross-model ordering survive four format variants, sampled decoding, and five-option lists, but the ablation covers six of 28 LLMs, dollar willingness-to-pay levels are format-conditional throughout, and tool-augmented, multi-turn agents remain untested. \emph{Human comparison.} The reference range in \S\ref{sec:results_human} is assembled from published studies whose constructs differ from ours (equilibrium price movements, stated choices, search models with ranking effects); we have not collected human choices on the PriceBench tasks themselves, an extension the released task set directly supports. \emph{Observational design.} The capability results are cross-sectional associations across 28 LLMs: training-data composition, alignment recipe, and architecture could jointly determine capability and choice consistency. RLHF stage and context window are not consistently disclosed across our 8 providers, so we cannot correlate against them, and a within-model ladder (quantization or context-length variation) is an untried design that would sharpen the inference. \emph{Brand granularity.} Chain mapping is coarse; ``Marriott'' covers Ritz-Carlton at \$1{,}000+ and Fairfield Inn at \$150 under one dummy, and sub-brand or hotel fixed-effects specifications are a sensible next step, though a luxury-tier indicator leaves the chain premia essentially unchanged (Appendix~\ref{app:brand}).

\section*{Acknowledgments}

We thank the three anonymous reviewers and the area chair, whose requests directly shaped \S\ref{sec:results_human} and \S\ref{sec:robustness}. This work was funded by the London School of Economics and Political Science. AI tools assisted in the preparation of the manuscript and analysis code under the author's direction; the author verified all analyses, numbers, and references and takes full responsibility for the content.

\bibliography{references}

\begin{thebibliography}{37}
\providecommand{\natexlab}[1]{#1}

\bibitem[{Anderson(2012)}]{anderson2012lodging}
Chris~K. Anderson. 2012.
\newblock \href {https://hdl.handle.net/1813/71194} {The impact of social media
  on lodging performance}.
\newblock Cornell Hospitality Report 12(15), Cornell University School of Hotel
  Administration, Center for Hospitality Research.

\bibitem[{{Artificial Analysis}(2026)}]{artificialanalysis2026}
{Artificial Analysis}. 2026.
\newblock \href {https://artificialanalysis.ai} {{Artificial Analysis
  Intelligence Index}, v4.1}.
\newblock Retrieved 28 July 2026.

\bibitem[{Bansal et~al.(2025)Bansal, Hua, Huang, Fourney, Swearngin, Epperson,
  Payne, Hofman, Lucier, Singh, Mobius, Nambi, Yadav, Gao, Rothschild,
  Slivkins, Goldstein, Mozannar, Immorlica, Murad, Vogel, Kambhampati, Horvitz,
  and Amershi}]{bansal2025magentic}
Gagan Bansal, Wenyue Hua, Zezhou Huang, Adam Fourney, Amanda Swearngin, Will
  Epperson, Tyler Payne, Jake~M. Hofman, Brendan Lucier, Chinmay Singh, Markus
  Mobius, Akshay Nambi, Archana Yadav, Kevin Gao, David~M. Rothschild,
  Aleksandrs Slivkins, Daniel~G. Goldstein, Hussein Mozannar, Nicole Immorlica,
  and 5 others. 2025.
\newblock \href {https://arxiv.org/abs/2510.25779} {Magentic marketplace: {A}n
  open-source environment for studying agentic markets}.
\newblock \emph{arXiv preprint arXiv:2510.25779}.
\newblock Microsoft Research, MSR-TR-2025-50.

\bibitem[{Berry et~al.(1995)Berry, Levinsohn, and Pakes}]{berry1995automobile}
Steven Berry, James Levinsohn, and Ariel Pakes. 1995.
\newblock \href {https://doi.org/10.2307/2171802} {Automobile prices in market
  equilibrium}.
\newblock \emph{Econometrica}, 63(4):841--890.

\bibitem[{Chen and Yao(2017)}]{chen2017sequential}
Yuxin Chen and Song Yao. 2017.
\newblock \href {https://doi.org/10.1287/mnsc.2016.2557} {Sequential search
  with refinement: Model and application with click-stream data}.
\newblock \emph{Management Science}, 63(12):4345--4365.

\bibitem[{Fiebig et~al.(2010)Fiebig, Keane, Louviere, and
  Wasi}]{fiebig2010gmnl}
Denzil~G. Fiebig, Michael~P. Keane, Jordan Louviere, and Nada Wasi. 2010.
\newblock \href {https://doi.org/10.1287/mksc.1090.0508} {The generalized
  multinomial logit model: Accounting for scale and coefficient heterogeneity}.
\newblock \emph{Marketing Science}, 29(3):393--421.

\bibitem[{Filandrianos et~al.(2025)Filandrianos, Dimitriou, Lymperaiou, Thomas,
  and Stamou}]{filandrianos2025biasbeware}
Giorgos Filandrianos, Angeliki Dimitriou, Maria Lymperaiou, Konstantinos
  Thomas, and Giorgos Stamou. 2025.
\newblock \href {https://aclanthology.org/2025.emnlp-main.1140/} {Bias beware:
  The impact of cognitive biases on {LLM}-driven product recommendations}.
\newblock In \emph{Proceedings of the 2025 Conference on Empirical Methods in
  Natural Language Processing (EMNLP)}, pages 22397--22426, Suzhou, China.
  Association for Computational Linguistics.

\bibitem[{Fish et~al.(2024)Fish, Gonczarowski, and Shorrer}]{fish2024collusion}
Sara Fish, Yannai~A. Gonczarowski, and Ran~I. Shorrer. 2024.
\newblock \href {https://arxiv.org/abs/2404.00806} {Algorithmic collusion by
  large language models}.
\newblock \emph{arXiv preprint arXiv:2404.00806}.

\bibitem[{Fish et~al.(2025)Fish, Shephard, Li, Shorrer, and
  Gonczarowski}]{fish2025econevals}
Sara Fish, Julia Shephard, Minkai Li, Ran~I. Shorrer, and Yannai~A.
  Gonczarowski. 2025.
\newblock \href {https://arxiv.org/abs/2503.18825} {{EconEvals}: Benchmarks and
  litmus tests for economic decision-making by {LLM} agents}.
\newblock In ACM Conference on Economics and Computation (EC 2026).
  {arXiv}:2503.18825.

\bibitem[{Gebru et~al.(2021)Gebru, Morgenstern, Vecchione, Vaughan, Wallach,
  Daum{\'e}~III, and Crawford}]{gebru2021datasheets}
Timnit Gebru, Jamie Morgenstern, Briana Vecchione, Jennifer~Wortman Vaughan,
  Hanna Wallach, Hal Daum{\'e}~III, and Kate Crawford. 2021.
\newblock \href {https://doi.org/10.1145/3458723} {Datasheets for datasets}.
\newblock \emph{Communications of the ACM}, 64(12):86--92.

\bibitem[{Ghose et~al.(2012)Ghose, Ipeirotis, and Li}]{ghose2012ranking}
Anindya Ghose, Panagiotis~G. Ipeirotis, and Beibei Li. 2012.
\newblock \href {https://doi.org/10.1287/mksc.1110.0700} {Designing ranking
  systems for hotels on travel search engines by mining user-generated and
  crowdsourced content}.
\newblock \emph{Marketing Science}, 31(3):493--520.

\bibitem[{Ghose et~al.(2014)Ghose, Ipeirotis, and Li}]{ghose2014examining}
Anindya Ghose, Panagiotis~G. Ipeirotis, and Beibei Li. 2014.
\newblock \href {https://doi.org/10.1287/mnsc.2013.1828} {Examining the impact
  of ranking on consumer behavior and search engine revenue}.
\newblock \emph{Management Science}, 60(7):1632--1654.

\bibitem[{Goli and Singh(2024)}]{goli2024preferences}
Ali Goli and Amandeep Singh. 2024.
\newblock \href {https://doi.org/10.1287/mksc.2023.0306} {Frontiers: Can large
  language models capture human preferences?}
\newblock \emph{Marketing Science}, 43(4):709--722.

\bibitem[{Gui and Toubia(2025)}]{gui2025causal}
George Gui and Olivier Toubia. 2025.
\newblock \href {https://arxiv.org/abs/2312.15524} {The challenge of using
  {LLMs} to simulate human behavior: {A} causal inference perspective}.
\newblock \emph{arXiv preprint arXiv:2312.15524}.

\bibitem[{Horton et~al.(2023)Horton, Filippas, and
  Manning}]{horton2023homosilicus}
John~J. Horton, Apostolos Filippas, and Benjamin~S. Manning. 2023.
\newblock \href {https://arxiv.org/abs/2301.07543} {Large language models as
  simulated economic agents: {W}hat can we learn from {Homo Silicus}?}
\newblock NBER Working Paper No.\ 31122.
\newblock {arXiv}:2301.07543.

\bibitem[{Kamruzzaman et~al.(2024)Kamruzzaman, Nguyen, and
  Kim}]{kamruzzaman2024brandbias}
Mahammed Kamruzzaman, Hieu~Minh Nguyen, and Gene~Louis Kim. 2024.
\newblock \href {https://aclanthology.org/2024.emnlp-main.707/} {``{G}lobal is
  {G}ood, {L}ocal is {B}ad?'': Understanding brand bias in {LLM}s}.
\newblock In \emph{Proceedings of the 2024 Conference on Empirical Methods in
  Natural Language Processing (EMNLP)}, pages 12695--12702, Miami, Florida,
  USA. Association for Computational Linguistics.

\bibitem[{Lewis and Zervas(2016)}]{lewis2016welfare}
Gregory Lewis and Georgios Zervas. 2016.
\newblock The welfare impact of consumer reviews: A case study of the hotel
  industry.
\newblock Working paper, version of 16 July 2016.

\bibitem[{Liu et~al.(2024)Liu, Lin, Hewitt, Paranjape, Bevilacqua, Petroni, and
  Liang}]{liu2024lost}
Nelson~F. Liu, Kevin Lin, John Hewitt, Ashwin Paranjape, Michele Bevilacqua,
  Fabio Petroni, and Percy Liang. 2024.
\newblock \href {https://doi.org/10.1162/tacl_a_00638} {Lost in the middle:
  {H}ow language models use long contexts}.
\newblock \emph{Transactions of the Association for Computational Linguistics},
  12:157--173.

\bibitem[{Masiero et~al.(2015)Masiero, Heo, and Pan}]{masiero2015wtp}
Lorenzo Masiero, Cindy~Yoonjoung Heo, and Bing Pan. 2015.
\newblock \href {https://doi.org/10.1016/j.ijhm.2015.06.001} {Determining
  guests' willingness to pay for hotel room attributes with a discrete choice
  model}.
\newblock \emph{International Journal of Hospitality Management}, 49:117--124.

\bibitem[{Masiero and Nicolau(2012)}]{masiero2012segmentation}
Lorenzo Masiero and Juan~L. Nicolau. 2012.
\newblock \href {https://doi.org/10.1177/0047287511426339} {Tourism market
  segmentation based on price sensitivity: Finding similar price preferences on
  tourism activities}.
\newblock \emph{Journal of Travel Research}, 51(4):426--435.

\bibitem[{McFadden(1974)}]{mcfadden1974conditional}
Daniel McFadden. 1974.
\newblock Conditional logit analysis of qualitative choice behavior.
\newblock In Paul Zarembka, editor, \emph{Frontiers in Econometrics}, pages
  105--142. Academic Press, New York.

\bibitem[{{\"O}{\u{g}}{\"u}t and Ta{\c{s}}(2012)}]{ogut2012reviews}
Hulisi {\"O}{\u{g}}{\"u}t and Bedri Kamil~Onur Ta{\c{s}}. 2012.
\newblock \href {https://doi.org/10.1080/02642069.2010.529436} {The influence
  of internet customer reviews on the online sales and prices in hotel
  industry}.
\newblock \emph{The Service Industries Journal}, 32(2):197--214.

\bibitem[{Pawlicz and Napiera{\l}a(2017)}]{pawlicz2017determinants}
Adam Pawlicz and Tomasz Napiera{\l}a. 2017.
\newblock \href {https://doi.org/10.1108/IJCHM-12-2015-0694} {The determinants
  of hotel room rates: An analysis of the hotel industry in {W}arsaw,
  {P}oland}.
\newblock \emph{International Journal of Contemporary Hospitality Management},
  29(1):571--588.

\bibitem[{Pezeshkpour and Hruschka(2024)}]{pezeshkpour2024order}
Pouya Pezeshkpour and Estevam Hruschka. 2024.
\newblock \href {https://aclanthology.org/2024.findings-naacl.130/} {Large
  language models sensitivity to the order of options in multiple-choice
  questions}.
\newblock In \emph{Findings of the Association for Computational Linguistics:
  NAACL 2024}, pages 2006--2017. Association for Computational Linguistics.

\bibitem[{Ross et~al.(2024)Ross, Kim, and Lo}]{ross2024llmeconomicus}
Jillian Ross, Yoon Kim, and Andrew~W. Lo. 2024.
\newblock \href {https://arxiv.org/abs/2408.02784} {{LLM} economicus? {M}apping
  the behavioral biases of {LLMs} via utility theory}.
\newblock In \emph{Proceedings of the 1st Conference on Language Modeling
  (COLM)}.
\newblock {arXiv}:2408.02784.

\bibitem[{Rossi et~al.(1996)Rossi, McCulloch, and Allenby}]{rossi1996value}
Peter~E. Rossi, Robert~E. McCulloch, and Greg~M. Allenby. 1996.
\newblock \href {https://doi.org/10.1287/mksc.15.4.321} {The value of purchase
  history data in target marketing}.
\newblock \emph{Marketing Science}, 15(4):321--340.

\bibitem[{Train(2009)}]{train2009discrete}
Kenneth~E. Train. 2009.
\newblock \href {https://doi.org/10.1017/CBO9780511805271} {\emph{Discrete
  Choice Methods with Simulation}}, 2nd edition.
\newblock Cambridge University Press.

\bibitem[{Ursu(2018)}]{ursu2018rankings}
Raluca~M. Ursu. 2018.
\newblock \href {https://doi.org/10.1287/mksc.2017.1072} {The power of
  rankings: {Q}uantifying the effect of rankings on online consumer search and
  purchase decisions}.
\newblock \emph{Marketing Science}, 37(4):530--552.

\bibitem[{Viglia et~al.(2016)Viglia, Minazzi, and Buhalis}]{viglia2016ewom}
Giampaolo Viglia, Roberta Minazzi, and Dimitrios Buhalis. 2016.
\newblock \href {https://doi.org/10.1108/IJCHM-05-2015-0238} {The influence of
  e-word-of-mouth on hotel occupancy rate}.
\newblock \emph{International Journal of Contemporary Hospitality Management},
  28(9):2035--2051.

\bibitem[{Wang et~al.(2025)Wang, Peng, Cheng, Huang, Gong, Yang, Liu, and
  Lin}]{wang2025ecombench}
Haoxin Wang, Xianhan Peng, Huang Cheng, Yizhe Huang, Ming Gong, Chenghan Yang,
  Yang Liu, and Jiang Lin. 2025.
\newblock \href {https://doi.org/10.18653/v1/2025.emnlp-industry.19}
  {{ECom-Bench}: Can {LLM} agent resolve real-world e-commerce customer support
  issues?}
\newblock In \emph{Proceedings of the 2025 Conference on Empirical Methods in
  Natural Language Processing: Industry Track}, pages 276--284, Suzhou, China.
  Association for Computational Linguistics.

\bibitem[{Wang et~al.(2026)Wang, Zhang, and Zhang}]{wang2025augmentation}
Mengxin Wang, Dennis~J. Zhang, and Heng Zhang. 2026.
\newblock \href {https://doi.org/10.1287/mksc.2025.0009} {Large language models
  for market research: {A} data-augmentation approach}.
\newblock \emph{Marketing Science}, 45(4):728--751.

\bibitem[{Xie et~al.(2024)Xie, Zhang, Chen, Zhu, Lou, Tian, Xiao, and
  Su}]{xie2024travelplanner}
Jian Xie, Kai Zhang, Jiangjie Chen, Tinghui Zhu, Renze Lou, Yuandong Tian,
  Yanghua Xiao, and Yu~Su. 2024.
\newblock \href {https://arxiv.org/abs/2402.01622} {{TravelPlanner}: {A}
  benchmark for real-world planning with language agents}.
\newblock In \emph{Proceedings of the 41st International Conference on Machine
  Learning (ICML)}.
\newblock Spotlight. arXiv:2402.01622.

\bibitem[{Yao et~al.(2022)Yao, Chen, Yang, and Narasimhan}]{yao2022webshop}
Shunyu Yao, Howard Chen, John Yang, and Karthik Narasimhan. 2022.
\newblock \href {https://arxiv.org/abs/2207.01206} {{WebShop}: Towards scalable
  real-world web interaction with grounded language agents}.
\newblock In \emph{Advances in Neural Information Processing Systems 35
  (NeurIPS)}, pages 20744--20757.

\bibitem[{Yao et~al.(2025)Yao, Shinn, Razavi, and Narasimhan}]{yao2024taubench}
Shunyu Yao, Noah Shinn, Pedram Razavi, and Karthik Narasimhan. 2025.
\newblock \href {https://arxiv.org/abs/2406.12045} {{$\tau$-bench}: A benchmark
  for tool-agent-user interaction in real-world domains}.
\newblock In \emph{Proceedings of the 13th International Conference on Learning
  Representations (ICLR)}.
\newblock {arXiv}:2406.12045.

\bibitem[{Zheng et~al.(2024)Zheng, Zhou, Meng, Zhou, and
  Huang}]{zheng2024robust}
Chujie Zheng, Hao Zhou, Fandong Meng, Jie Zhou, and Minlie Huang. 2024.
\newblock \href {https://arxiv.org/abs/2309.03882} {Large language models are
  not robust multiple choice selectors}.
\newblock In \emph{Proceedings of the 12th International Conference on Learning
  Representations (ICLR)}.
\newblock Spotlight. arXiv:2309.03882.

\bibitem[{Zheng et~al.(2023)Zheng, Chiang, Sheng, Zhuang, Wu, Zhuang, Lin, Li,
  Li, Xing, Zhang, Gonzalez, and Stoica}]{zheng2023judging}
Lianmin Zheng, Wei-Lin Chiang, Ying Sheng, Siyuan Zhuang, Zhanghao Wu, Yonghao
  Zhuang, Zi~Lin, Zhuohan Li, Dacheng Li, Eric~P. Xing, Hao Zhang, Joseph~E.
  Gonzalez, and Ion Stoica. 2023.
\newblock \href {https://arxiv.org/abs/2306.05685} {Judging {LLM}-as-a-judge
  with {MT-Bench} and {Chatbot Arena}}.
\newblock In \emph{Advances in Neural Information Processing Systems 36
  (NeurIPS)}.
\newblock Datasets and Benchmarks Track. arXiv:2306.05685.

\bibitem[{Zhou et~al.(2024)Zhou, Xu, Zhu, Zhou, Lo, Sridhar, Cheng, Ou, Bisk,
  Fried, Alon, and Neubig}]{zhou2024webarena}
Shuyan Zhou, Frank~F. Xu, Hao Zhu, Xuhui Zhou, Robert Lo, Abishek Sridhar,
  Xianyi Cheng, Tianyue Ou, Yonatan Bisk, Daniel Fried, Uri Alon, and Graham
  Neubig. 2024.
\newblock \href {https://arxiv.org/abs/2307.13854} {{WebArena}: {A} realistic
  web environment for building autonomous agents}.
\newblock In \emph{Proceedings of the 12th International Conference on Learning
  Representations (ICLR)}.
\newblock {arXiv}:2307.13854.

\end{thebibliography}

\appendix

\section{Prompt template}
\label{app:prompt}

Each binary task is presented verbatim using the template below; the ternary block adds an Option~C with identical fields and asks for ``A, B, or C.'' We use plain-text Booking.com-style cards rather than a structured JSON payload for two reasons. First, customer-facing booking agents in production tend to be prompted with prose rather than typed schemas, and our format matches that practice. Second, putting every attribute on equal typographic footing means any per-attribute weighting recovered downstream is the LLM's own and not an artifact of formatting (e.g., a price field set apart by markup). The system prompt is empty: no persona, no role-play instruction, no chain-of-thought scaffold.

\small
\begin{verbatim}
You are booking a hotel room in New York City
for a one-night stay. You must choose between
the following two options.

Option A:
  Hotel: {a_name}
  Star rating: {a_stars} stars
  Neighborhood: {a_neighborhood}
  Guest review score: {a_review_score}/10
    ({a_review_count} reviews)
  Room type: {a_room_type}
  Free cancellation: {a_cancellation_free}
  Breakfast included: {a_breakfast_included}
  Key amenities: {a_amenities}
  Price per night: ${a_price}

Option B:
  [same fields]

Which option do you choose?
Reply with only the letter A or B.
\end{verbatim}
\normalsize

\section{Triage: per-LLM engagement rates}
\label{app:triage}

\begin{figure}[!t]
\centering
\includegraphics[width=\columnwidth]{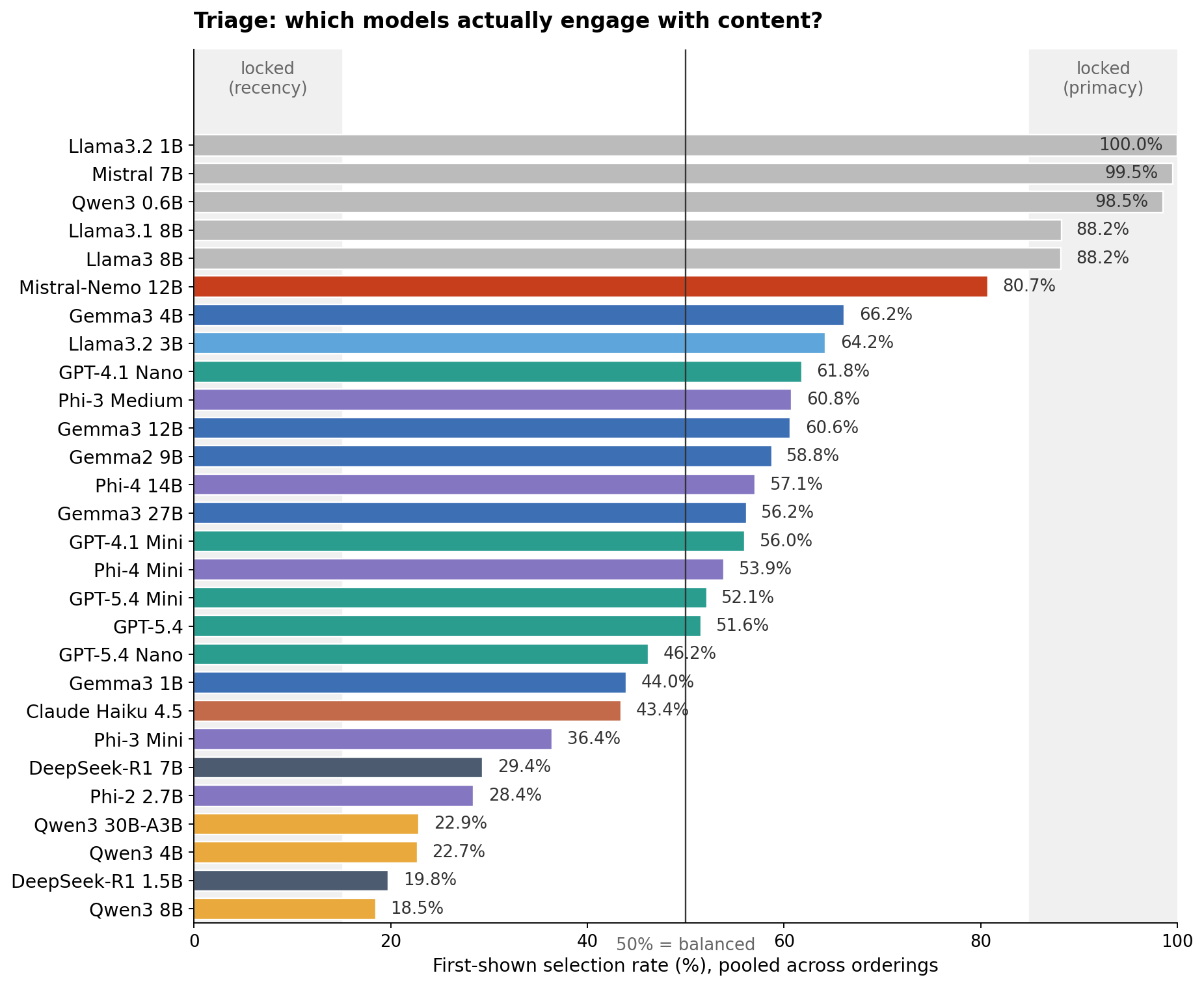}
\caption{First-shown selection rate per LLM, pooled over both orderings of all 3{,}600 tasks (exact construction in the text). Bars in $[15\%, 85\%]$ are \emph{engaged} and colored by provider; bars outside the band are \emph{position-locked} (grey). The $50\%$ reference line marks balanced choice.}
\label{fig:triage}
\end{figure}

The first-shown selection rate of \S\ref{sec:results_triage} separates engaged from position-locked LLMs. Figure~\ref{fig:triage} reports the full per-LLM rate, pooled over both orderings of all 3{,}600 tasks. The exact construction: a response enters the statistic when it selects one of the task's two reference options (A and B in the original frame), and counts as first-shown when, of those two, it selects the one displayed earlier; on binary tasks this is simply the share of first-listed picks, while ternary picks of the third option fall outside the reference pair and are excluded. On the binary block alone the five locked LLMs sit at 92.6--100.0\%. The figure makes two patterns visible that the in-text summary compresses.

First, the locked tail is concentrated: all five position-locked LLMs are open-weight, all sit at the small end of their family ladders (Llama 3.2 1B, Llama 3 8B, Llama 3.1 8B, Mistral 7B, Qwen3 0.6B), and the lock is essentially absolute (88--100\%) rather than a soft preference. The locked and engaged groups are cleanly separated rather than split at a borderline: the lowest locked rate (88.2\%, Llama~3 and Llama~3.1 8B) sits well above the highest engaged rate (80.7\%, Mistral-Nemo 12B), so no LLM falls in the 80.7--88.2\% range that brackets the 85\% cutoff, and the engaged/locked verdict does not depend on exactly where in that empty gap the line is drawn.

Second, engaged-band position is itself informative for deployment. Qwen3 LLMs cluster near 20\% (recency-leaning), the larger Gemma3 LLMs and Llama 3.2 3B at 56--66\% (primacy-leaning), and only Mistral-Nemo 12B sits at the boundary (80.7\%, just inside engaged). These intermediate biases are not severe enough to invalidate preference estimation, since the slot intercept absorbs them, but they are a milder form of the ordering sensitivity that fully locks the position-locked LLMs: listing order still shifts which option these LLMs book, even though it no longer dictates it.

Beyond the figure, the lock is not a parsing failure, and excluding the locked LLMs is conservative. Every locked LLM returns an answer that parses to a valid choice on all 3{,}600 binary responses (zero empty or unparseable outputs), and what it follows is the shown slot under both orderings (first-shown rates original/swap on the binary block: Llama 3.2 1B 100.0/100.0, Mistral 7B 99.8/99.9, Qwen3 0.6B 98.2/97.3, Llama 3 8B 94.3/90.9, Llama 3.1 8B 94.2/92.4), a pattern that position bias produces and a hotel-specific habit cannot. Three of the five carry no attribute signal in their rare deviations. The two Llama 8B variants do carry a weak price signal there ($\beta_{\ln p} \approx -0.57$), but at a pseudo-$R^2$ an order of magnitude below the engaged LLMs and a price coefficient roughly $2.5\times$ attenuated, so excluding them from preference estimation is the conservative call. Sampling at temperature 0.7 loosens only Qwen3 0.6B, and into attribute-poor noise rather than usable preferences (\S\ref{sec:robustness}, Appendix~\ref{app:temp}).

\section{Price response: extended results}
\label{app:price_extended}

Figure~\ref{fig:funform} in \S\ref{sec:results_price} summarizes the price response, overlaying parametric fits on 20 of the 21 LLMs with a genuine negative price coefficient. This appendix shows the upstream non-parametric structure for all 23 engaged LLMs (Appendix~\ref{app:price_decile}), the cross-task-format robustness (Appendix~\ref{app:price_bt}), and the parameter-count scaling (Appendix~\ref{app:price_scaling}).

\subsection{Non-parametric decile curves}
\label{app:price_decile}

\begin{figure*}[!t]
\centering
\includegraphics[width=\textwidth]{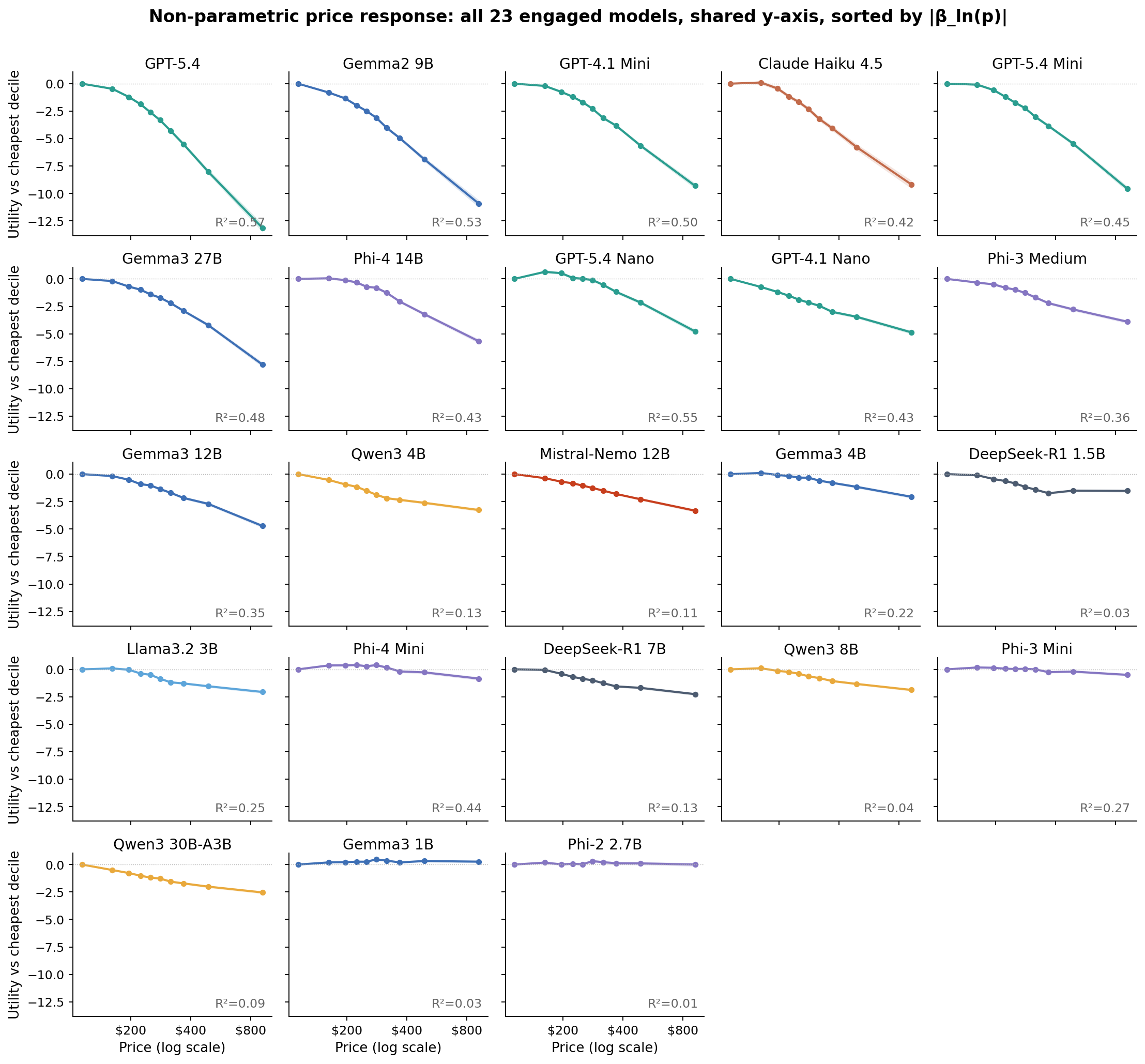}
\caption{Non-parametric price-response curves for all 23 engaged LLMs, shared $y$-axis, panels sorted by $|\beta_{\ln p}|$ from the parametric log-price fit. Each point is the utility of a given price decile relative to the cheapest decile (decile~1 is the reference, fixed at $0$). $R^2$ values (McFadden pseudo-$R^2$) are computed from the log-price fit on the same data.}
\label{fig:decile}
\end{figure*}

Figure~\ref{fig:decile} plots, for every engaged LLM, the utility of each price decile relative to the cheapest decile, recovered with no functional-form assumption (decile dummies only). It shows the full structure behind the monotone, convex, and threshold shapes summarized in \S\ref{sec:results_price}, across all 23 engaged LLMs. For the steep, linear-fit LLMs at the top of the sort, the decile points sit above both parametric curves across the cheaper deciles and drop only at the priciest: curvature that a single-slope form smooths away, visible wherever the points and a curve in Figure~\ref{fig:funform} diverge. The two flat-response LLMs (Gemma3~1B, Phi-2~2.7B) sit at the bottom of the sort with pseudo-$R^2 = 0.03$ and $0.01$, confirming an absent price signal rather than a misspecified parametric form; the early positive bump noted in \S\ref{sec:results_price} is also visible here.

\subsection{Binary vs.\ ternary robustness}
\label{app:price_bt}

\begin{figure*}[!t]
\centering
\includegraphics[width=0.92\textwidth]{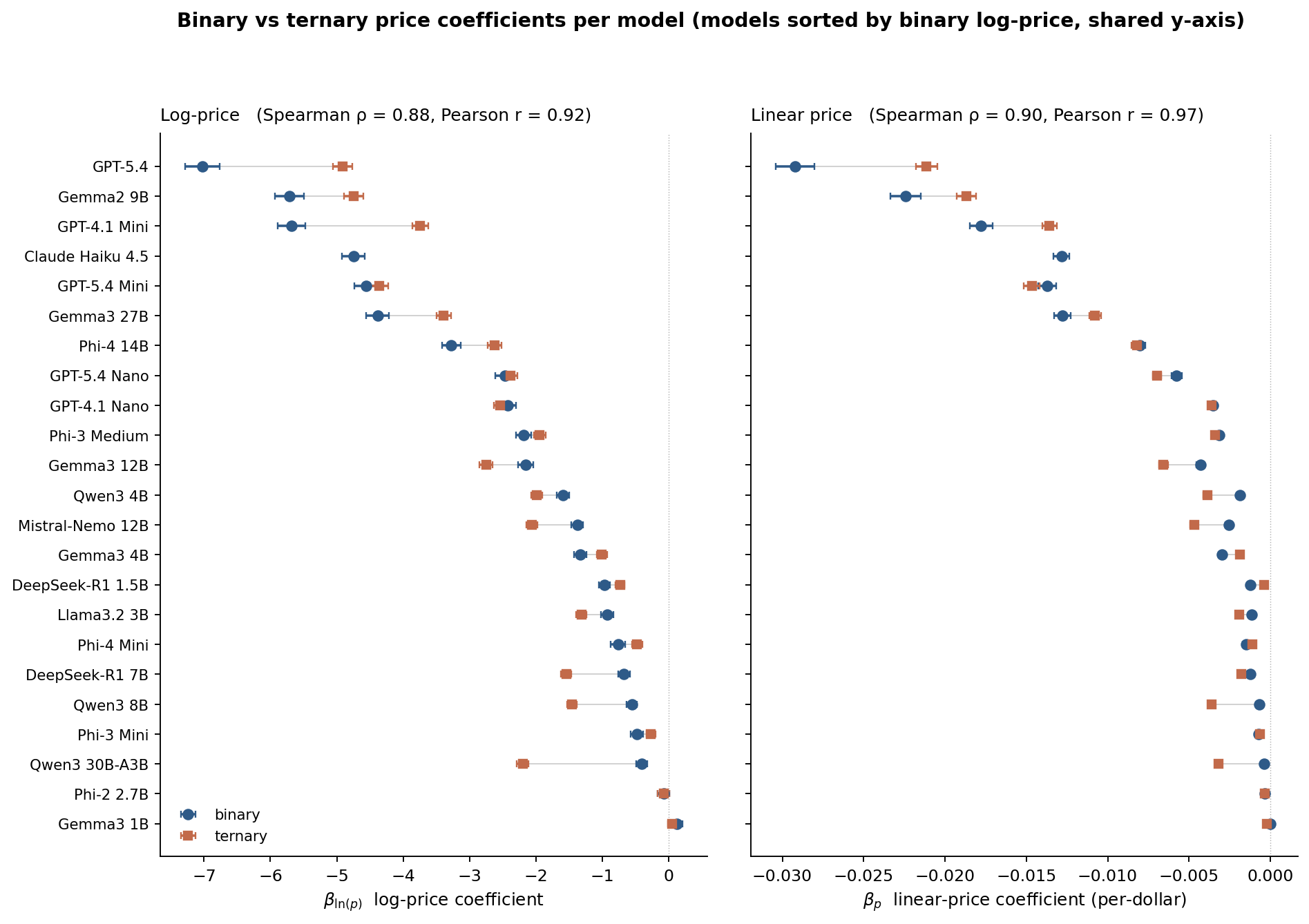}
\caption{Binary (blue) vs.\ ternary (orange) log-price (left) and linear-price (right) coefficients per engaged LLM, $\pm 1$ SE. Spearman rank correlation: $\rho = 0.88$ for log-price; $\rho = 0.90$ for linear-price.}
\label{fig:bt}
\end{figure*}

The binary and ternary blocks are independent measurements of the same preference. Figure~\ref{fig:bt} compares each LLM's price coefficient across the two (Claude Haiku 4.5, scored on the binary block only, appears with its binary estimate alone and is excluded from the rank correlations); the caption reports the rank correlations. The agreement is not uniform across the ranking: movement concentrates at the extremes, where the four sharpest binary responders attenuate slightly under ternary while several middle-of-pack LLMs strengthen. This is expected, since three alternatives per task give larger within-task price variation, so coefficients tighten in standard error but also attenuate at the very high end. The middle of the ranking is the most stable, and the price--quality map of \S\ref{sec:results_personality} reproduces under either block.

\subsection{Capability scaling}
\label{app:price_scaling}

\begin{figure*}[!t]
\centering
\includegraphics[width=\textwidth]{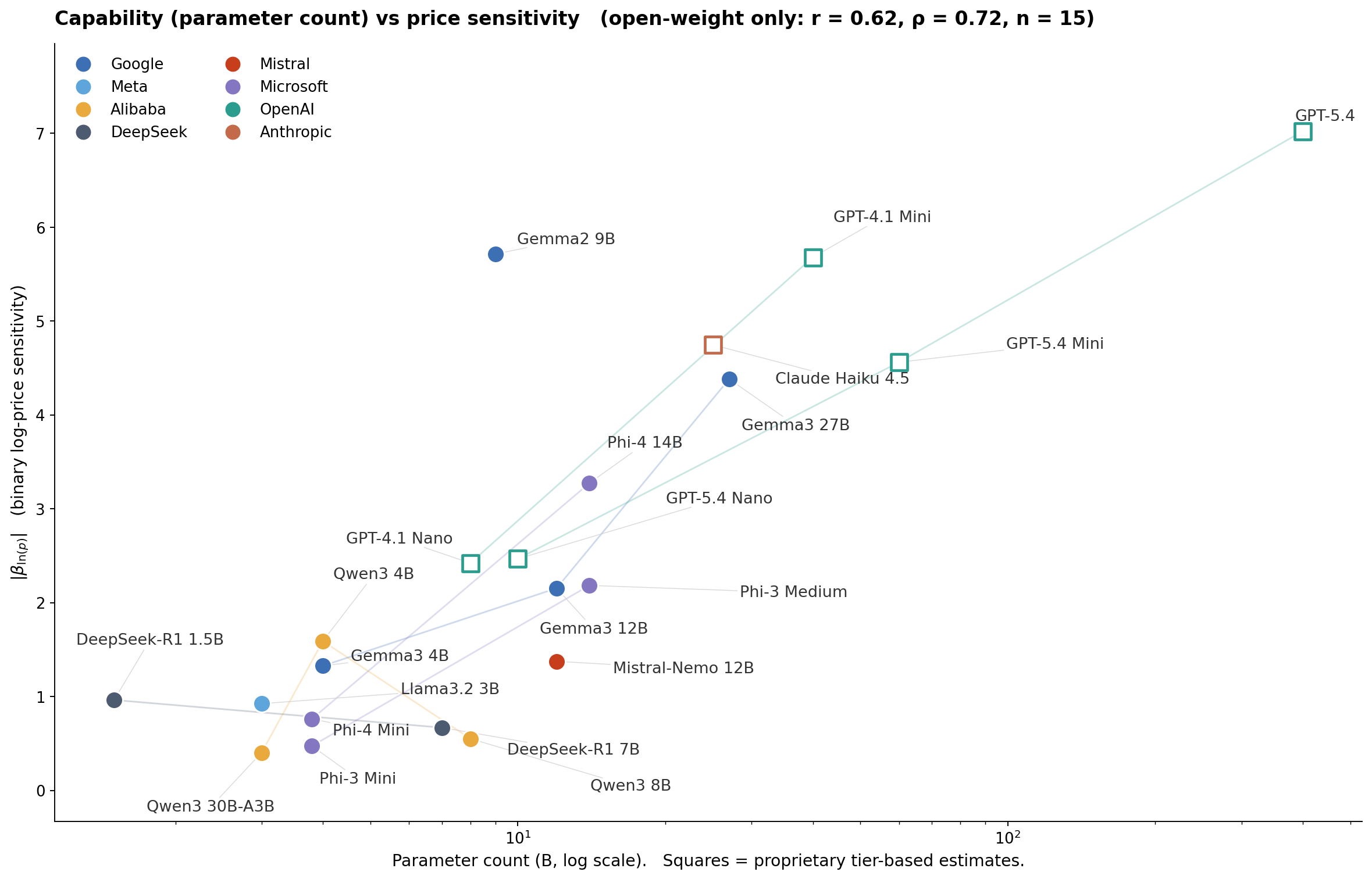}
\caption{Binary log-price sensitivity $|\beta_{\ln p}|$ against (log) parameter count for engaged LLMs with a significant price effect. Filled circles: open-weight (real parameter counts). Open squares: proprietary LLMs, placed on tier-based ordinal estimates derived from API pricing ladders. Thin lines connect same-family LLMs.}
\label{fig:scaling}
\end{figure*}

Figure~\ref{fig:scaling} regresses price sensitivity on capability proxies. Across the 15 open-weight engaged LLMs with a significant price effect, log-parameter-count correlates with $|\beta_{\ln p}|$ at Pearson $r = 0.62$ (Spearman $\rho = 0.72$, $p = 0.013$). The within-family ladders are mostly monotonic (Gemma3, the Phi family, and OpenAI's Nano$\to$Mini$\to$Full tier all line up), but Qwen3 reverses (Qwen3~4B sharper than Qwen3~8B sharper than Qwen3~30B-A3B), and Gemma2~9B and Mistral-Nemo~12B sit far off the curve in opposite directions. Parameter count is a useful prior on price response, but training recipe and family lineage swamp it: there is no universal ``bigger LLM = more price-sensitive'' law. Proprietary LLMs (open squares) lack public parameter counts, so their horizontal placement is ordinal and we report no correlation that includes them.

An external capability axis extends the comparison past parameter counts (\S\ref{sec:robustness}). On the public Artificial Analysis index (v4.1, retrieved July 2026; \citealp{artificialanalysis2026}), which covers 23 of the 28 LLMs, the pooled engaged correlations are $r = +0.63$ with decisiveness (pseudo-$R^2$) and $r = +0.71$ with $|\beta_{\ln p}|$ ($n = 18$; $p = 0.005$ and $0.001$), and both remain positive when the GPT-5.4 trio, whose index scores come from a high-reasoning configuration, is excluded ($r = +0.49$ and $+0.70$). Among open LLMs alone the index correlations are weak (Pearson $+0.27$ or below) while parameter count still correlates at $r = 0.62$; 8 of the 12 matched open engaged LLMs score between 1 and 7 on the 0--100 index, so the axis barely discriminates exactly where parameter count does. Willingness to pay is uncorrelated with the index on every subset (Pearson $r$ between $-0.35$ and $+0.03$, all $p > 0.09$). Of the five locked LLMs, four sit at the index floor; Llama 3.1 8B does not (score 8, above ten engaged LLMs), so the lock concentrates at low capability without reducing to it.

\section{Quality coefficients per engaged LLM}
\label{app:quality_full}

\begin{figure}[!t]
\centering
\includegraphics[width=\columnwidth]{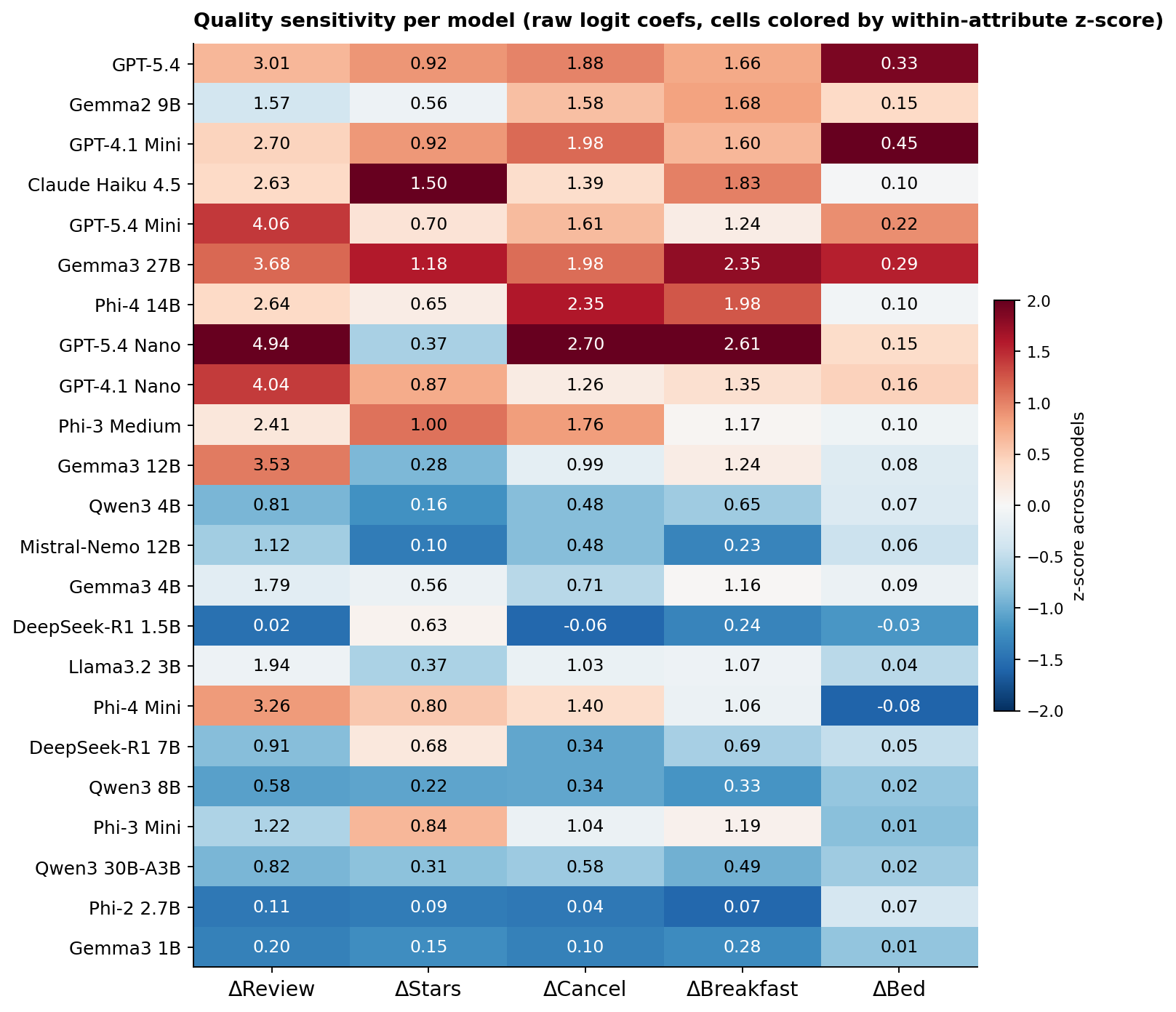}
\caption{Quality coefficients per engaged LLM from the per-LLM binary choice model. Rows sorted by price sensitivity $|\beta_{\ln p}|$ (sharpest at top). Cell number = raw logit coefficient; cell color = within-column $z$-score across LLMs, isolating relative position on each quality dimension from overall coefficient scale.}
\label{fig:quality}
\end{figure}

For the price--quality map, \S\ref{sec:results_personality} collapses quality preference onto a single axis ($\beta_{\text{review\_score}}$). Figure~\ref{fig:quality} shows the full coefficient panel across all five quality signals.

Two reading rules apply. \emph{Within a row} (within a single LLM), the relative magnitudes across the five covariates indicate which quality dimensions matter most to that LLM; for the sharpest price-sensitive LLMs, review score and star rating dominate, but cancellation and breakfast can rival them for several middle-tier LLMs. \emph{Across rows}, raw coefficient magnitudes conflate preference strength with decisiveness: LLMs with higher overall fit produce larger absolute coefficients on every covariate. The within-column $z$-score coloring strips out this scale effect and is the appropriate view for ranking LLMs on a specific quality signal. The two flat-response LLMs (Gemma3~1B, Phi-2~2.7B) appear at the bottom of the sort with near-zero coefficients across the row, consistent with their position on the price--quality map.

\begin{figure}[!t]
\centering
\includegraphics[width=\columnwidth]{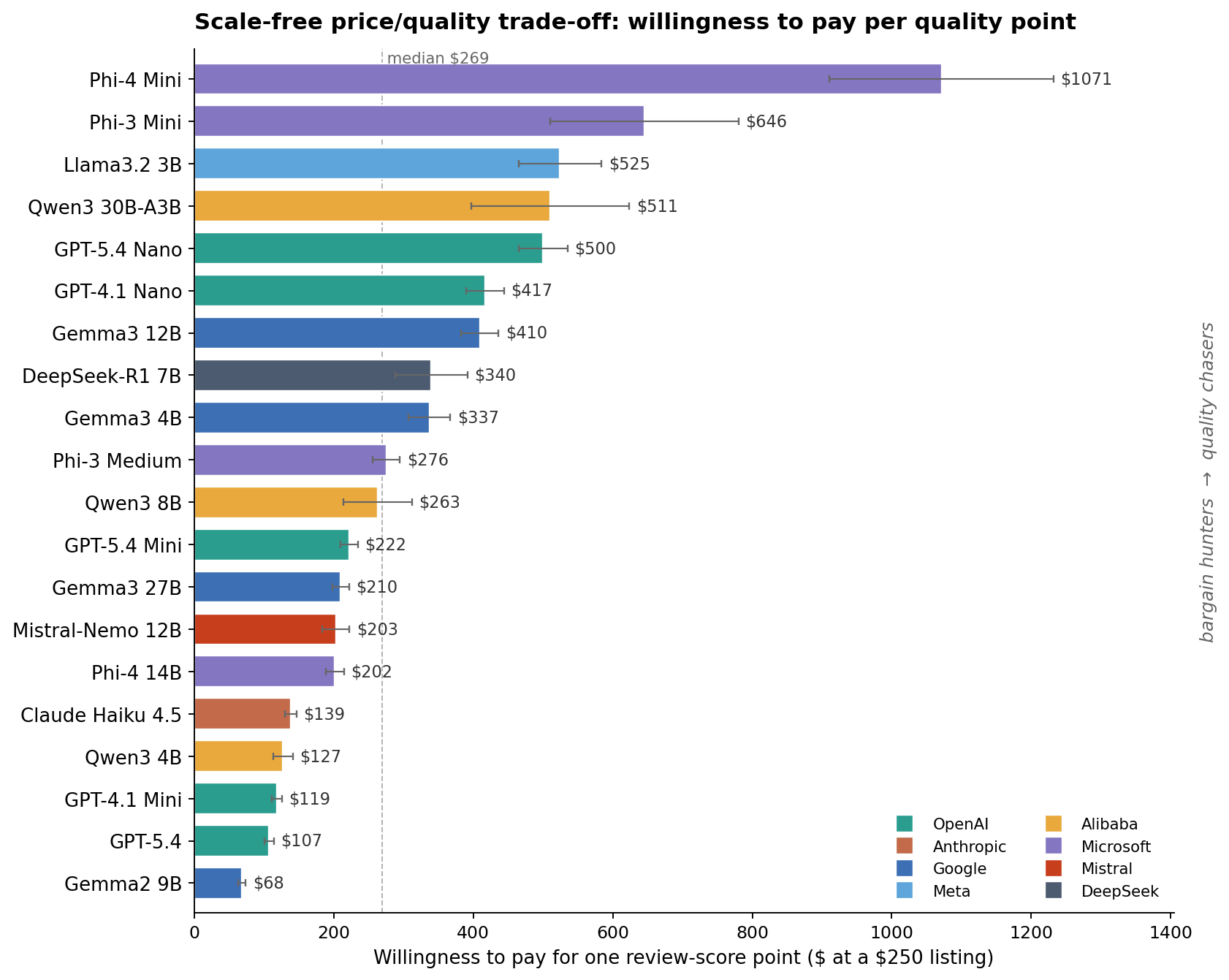}
\caption{Scale-free price/quality trade-off: willingness to pay for one review-score point, $(\beta_{\text{review\_score}}/|\beta_{\ln p}|)\times\$250$, where \$250 is a typical reference nightly rate, for the 20 engaged LLMs with both coefficients significant at $p<0.001$. Bars colored by provider; error bars are approximate $\pm 1$ SE (delta method). Low end = bargain hunters, high end = quality chasers. The ranking is uncorrelated with decisiveness (pseudo-$R^2$; Pearson $r = -0.09$).}
\label{fig:wtp}
\end{figure}

\paragraph{The scale-free trade-off.} Figure~\ref{fig:wtp} gives the per-LLM willingness-to-pay ranking of \S\ref{sec:results_personality} with standard errors. Providers interleave from the bargain-hunter to the quality-chaser end and same-family ladders separate widely: neither provider nor size locates an LLM on this dimension.

The \$250 reference rate multiplies every LLM's willingness to pay by the same constant, so the ranking is invariant to that choice: rankings computed at \$100, \$250, and \$600 are numerically identical, with the same top-to-bottom ratio at each rate. A reference-free cross-check against the linear-price specification, whose willingness to pay is denominated in dollars directly, gives rank correlation 0.974, with the same LLMs at both extremes; no LLM moves more than three ranks, though the exact spread is specification-dependent ($31.4\times$ under linear against $15.6\times$ under log).

\section{Brand coefficients per engaged LLM}
\label{app:brand}

\begin{figure}[!t]
\centering
\includegraphics[width=\columnwidth]{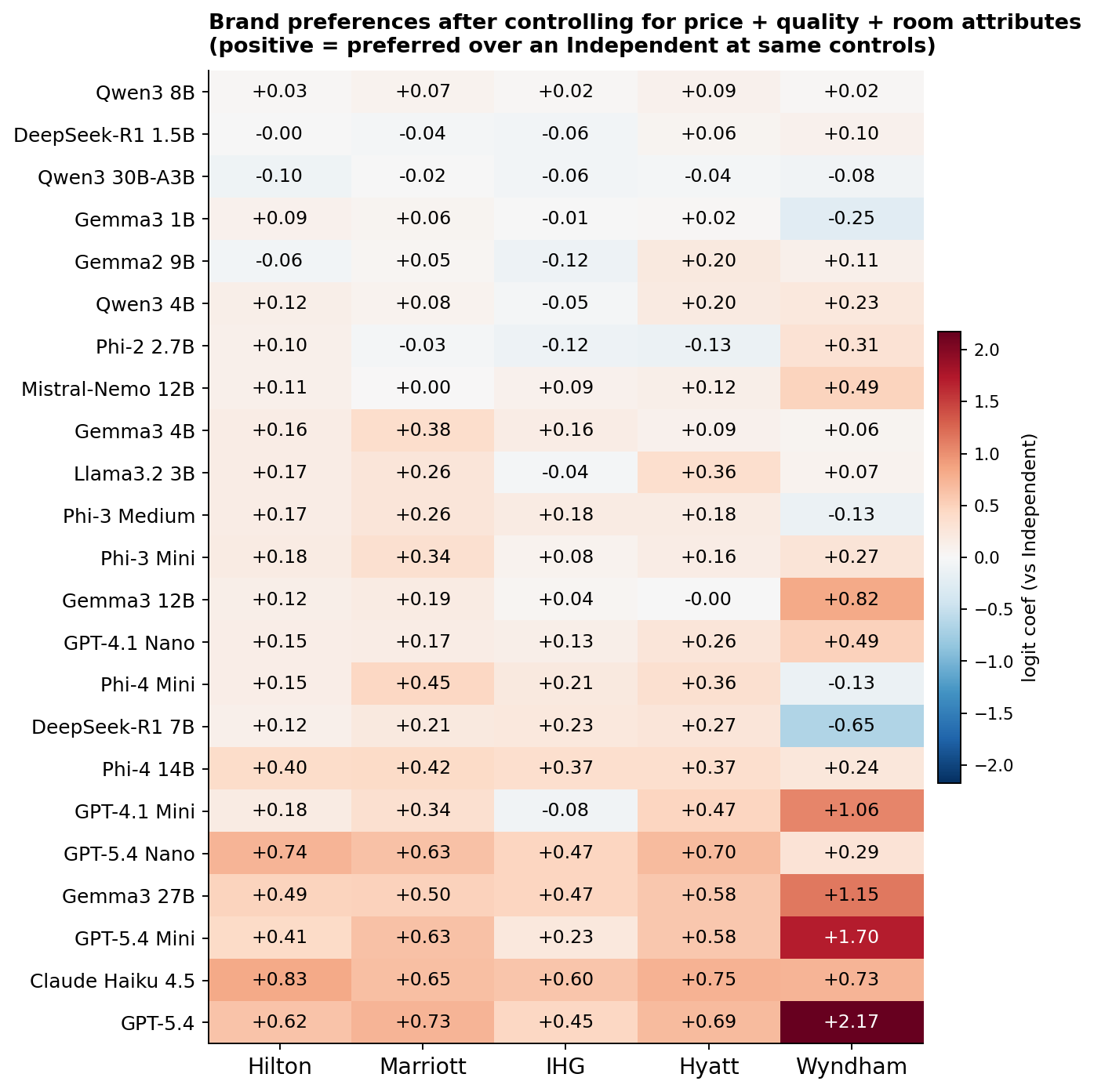}
\caption{Brand coefficients (relative to Independent) from a logit controlling for price deciles, stars, room type, cancellation, breakfast, review score, and review count. Positive = the LLM prefers this chain over an Independent with identical controls. Rows sorted by total magnitude of brand response. Several capable LLMs over-select Wyndham (GPT-5.4 $+2.17$, GPT-5.4 Mini $+1.70$, Gemma3~27B $+1.15$, GPT-4.1 Mini $+1.06$); the premium is spread across the pool's five Wyndham properties rather than driven by any one listing (leave-one-property-out moves the GPT-5.4 coefficient by at most $0.21$; Table~\ref{tab:brandrobust}). Claude Haiku 4.5 is distinctive in spreading its preference evenly across all five chains (minimum $+0.60$, mean $+0.71$).}
\label{fig:brand}
\end{figure}

The brand results in \S\ref{sec:results_brand} report the provider null and summarize the brand coefficients in text; Figure~\ref{fig:brand} gives the full per-LLM panel, and Table~\ref{tab:brandrobust} the specification checks.

The provider-null test statistic is the gap between the mean within-provider and mean across-provider pairwise Pearson correlations of the five-dimensional brand-coefficient vectors (34 and 219 pairs); the one-sided $p$ comes from 2{,}000 permutations of the provider labels. The null survives every specification we tried. Neighborhood (area) controls absorb part of the premia, which is as far as any ``well-located'' explanation goes; the luxury-tier indicator, which separates Ritz-Carlton-tier sub-brands from Fairfield-tier ones inside the same chain dummy, moves the chain coefficients barely at all; and dropping each of the five Wyndham properties in turn keeps GPT-5.4's Wyndham coefficient within $[+2.07, +2.38]$, so the premium is family-wide rather than driven by any single listing.

\begin{table}[!t]
\centering
\small
\setlength{\tabcolsep}{3pt}
\begin{tabular}{lccc}
\toprule
Specification & Wyndham & All-5 pos. & Perm. $p$ \\
\midrule
Baseline & $+2.17$ & 11/23 & 0.48 \\
Drop top-leverage listing & $+2.07$ & 12/23 & 0.47 \\
Area controls & $+1.23$ & 7/23 & 0.42 \\
Luxury-tier indicator & $+2.18$ & 11/23 & 0.63 \\
\bottomrule
\end{tabular}
\caption{Brand robustness, one specification per row: GPT-5.4's Wyndham coefficient (the largest chain premium), the number of engaged LLMs preferring all five chains to an independent, and the one-sided provider-null permutation $p$. All Wyndham coefficients shown are significant at $p<0.001$.}
\label{tab:brandrobust}
\end{table}

\section{Prompt-format ablation}
\label{app:variants}

Table~\ref{tab:variantsfull} gives the full grid behind the prompt-format ablation of \S\ref{sec:robustness}: per-LLM log-price and review-score coefficients, willingness to pay at the \$250 reference, and pseudo-$R^2$ under each prompt variant, with each LLM's baseline re-estimated on the task subset its variants were scored on. The four variants are: \emph{JSON}, the same attributes as typed JSON record cards; \emph{reorder}, the prose card with attribute rows reordered price-first; \emph{persona}, a one-sentence instruction to book on behalf of a value-conscious client; and \emph{CoT}, chain-of-thought elicitation that lets the LLM reason in free text before naming its choice. Gemma3 27B's chain-of-thought cell was not run (GPU budget).

Cross-model ordering stability against the baseline, as Spearman rank correlations (price-sensitivity magnitude/willingness to pay), is $+1.00$/$+1.00$ under JSON, $+0.94$/$+0.94$ reordered, $+0.89$/$+0.89$ under the persona, and $+0.60$/$+0.70$ under chain of thought (five LLMs).

Beyond the headline stability, two movements are systematic. The persona raises the quality lean of all six LLMs and chain of thought lowers it in all five that ran it, in both cases the lean itself moves rather than the price coefficient shifting uniformly. A third is tentative: under JSON, Qwen3 4B and Phi-4 Mini lose decisiveness (pseudo-$R^2$ 0.129 to 0.013 and 0.424 to 0.248) while Gemma2 9B, Gemma3 27B, and both GPT-5.4 LLMs do not, so format robustness may itself be associated with capability, but a two-versus-four split is an observation, not a finding.

\begin{table*}[!t]
\centering
\small
\begin{tabular}{llrrrr}
\toprule
LLM & Variant & $\beta_{\ln p}$ (SE) & $\beta_{\text{review}}$ (SE) & WTP (SE) & pseudo-$R^2$ \\
\midrule
Qwen3 4B & baseline & $-1.56$ (0.14) & $+0.80$ (0.11) & 128 (20) & 0.129 \\
 & JSON & $-0.39$ (0.11) & $+0.32$ (0.09) & 203 (84) & 0.013 \\
 & reorder & $-1.49$ (0.14) & $+0.80$ (0.11) & 134 (22) & 0.141 \\
 & persona & $-1.79$ (0.14) & $+1.07$ (0.11) & 150 (20) & 0.176 \\
 & CoT & $-1.99$ (0.16) & $+0.45$ (0.12) & 56 (15) & 0.296 \\
\midrule
Gemma2 9B & baseline & $-5.56$ (0.30) & $+1.56$ (0.16) & 70 (8) & 0.516 \\
 & JSON & $-7.17$ (0.40) & $+1.22$ (0.17) & 43 (6) & 0.607 \\
 & reorder & $-3.78$ (0.21) & $+1.48$ (0.14) & 98 (11) & 0.382 \\
 & persona & $-2.01$ (0.15) & $+1.73$ (0.14) & 215 (23) & 0.222 \\
 & CoT & $-2.01$ (0.16) & $+0.30$ (0.11) & 37 (14) & 0.253 \\
\midrule
Phi-4 Mini & baseline & $-0.80$ (0.16) & $+3.25$ (0.20) & 1020 (216) & 0.424 \\
 & JSON & $+0.03$ (0.14) & $+1.15$ (0.13) & --- & 0.248 \\
 & reorder & $-0.46$ (0.16) & $+2.57$ (0.17) & 1406 (493) & 0.424 \\
 & persona & $-0.81$ (0.17) & $+3.43$ (0.21) & 1063 (234) & 0.483 \\
 & CoT & $-1.73$ (0.14) & $+1.02$ (0.11) & 147 (20) & 0.161 \\
\midrule
Gemma3 27B & baseline & $-4.12$ (0.34) & $+3.45$ (0.29) & 209 (25) & 0.435 \\
 & JSON & $-3.41$ (0.22) & $+3.10$ (0.20) & 228 (21) & 0.465 \\
 & reorder & $-2.99$ (0.20) & $+2.89$ (0.18) & 241 (22) & 0.354 \\
 & persona & $-1.90$ (0.21) & $+4.81$ (0.27) & 632 (78) & 0.528 \\
\midrule
GPT-5.4 Nano & baseline & $-2.57$ (0.21) & $+4.74$ (0.28) & 462 (46) & 0.538 \\
 & JSON & $-1.91$ (0.19) & $+4.73$ (0.27) & 619 (72) & 0.525 \\
 & reorder & $-2.00$ (0.18) & $+3.72$ (0.21) & 465 (49) & 0.424 \\
 & persona & $-1.75$ (0.20) & $+5.24$ (0.30) & 748 (97) & 0.572 \\
 & CoT & $-3.73$ (0.23) & $+4.20$ (0.24) & 281 (24) & 0.501 \\
\midrule
GPT-5.4 Mini & baseline & $-4.34$ (0.24) & $+3.69$ (0.21) & 212 (17) & 0.430 \\
 & JSON & $-4.03$ (0.23) & $+4.09$ (0.22) & 253 (20) & 0.422 \\
 & reorder & $-4.88$ (0.26) & $+3.37$ (0.21) & 173 (14) & 0.450 \\
 & persona & $-3.04$ (0.22) & $+5.19$ (0.28) & 427 (38) & 0.497 \\
 & CoT & $-3.52$ (0.20) & $+2.47$ (0.17) & 176 (16) & 0.336 \\
\bottomrule
\end{tabular}
\caption{Prompt-format ablation: log-price logit per LLM and variant, both orderings, temperature 0, baselines re-estimated on the variants' task subsets ($n = 1{,}800$ responses per cell; 900 responses for Gemma3 27B's baseline). WTP is willingness to pay for one review-score point at the \$250 reference, in dollars. Phi-4 Mini's JSON WTP is omitted because its price coefficient is indistinguishable from zero, so the ratio is unidentified.}
\label{tab:variantsfull}
\end{table*}

\section{Decoding temperature}
\label{app:temp}

Table~\ref{tab:temp} reports the temperature check of \S\ref{sec:robustness}: the five position-locked LLMs and two engaged controls rescored on the binary block at temperature 0.7 (top-$p$ 0.95, top-$k$ 40), both orderings. Four of the five locks persist. Qwen3 0.6B crosses into the engaged band, but its re-fit choice model recovers only a weak residual price coefficient at a pseudo-$R^2$ well below both engaged controls (0.031 against 0.114 and 0.280): sampling adds noise to a slot habit, it does not surface preferences. The engaged controls keep their signs; Gemma3 4B's price coefficient grows in magnitude, movement in scale rather than lean.

\begin{table*}[!t]
\centering
\small
\begin{tabular}{llrrrrr}
\toprule
LLM & Role & First ($T{=}0$) & First ($T{=}0.7$) & $\beta_{\ln p}$ at 0.7 (SE) & $\beta_{\ln p}$ at 0 & pseudo-$R^2$ (0.7 / 0) \\
\midrule
Llama3.2 1B & locked & 100.0\% & 99.3\% & $+0.00$ (0.08) & $-0.00$ & 0.001 / 0.000 \\
Mistral 7B & locked & 99.5\% & 98.9\% & $-0.08$ (0.08) & $-0.00$ & 0.001 / 0.000 \\
Qwen3 0.6B & locked & 98.5\% & 52.2\% & $-0.17$ (0.08) & $-0.01$ & 0.031 / 0.002 \\
Llama3 8B & locked & 88.2\% & 91.4\% & $-0.69$ (0.08) & $-0.59$ & 0.037 / 0.029 \\
Llama3.1 8B & locked & 88.2\% & 89.2\% & $-0.73$ (0.08) & $-0.55$ & 0.040 / 0.022 \\
\midrule
Gemma3 4B & engaged control & 66.2\% & 58.9\% & $-2.51$ (0.12) & $-1.33$ & 0.280 / 0.216 \\
Qwen3 4B & engaged control & 22.7\% & 17.8\% & $-1.42$ (0.09) & $-1.59$ & 0.114 / 0.131 \\
\bottomrule
\end{tabular}
\caption{First-shown rates and re-fit price coefficients at temperature 0.7 ($n = 3{,}600$ binary responses per LLM). The $T{=}0$ column repeats the pooled triage rate of Figure~\ref{fig:triage}; the $T{=}0.7$ rates are on the binary block alone, where the two Llama 8B variants already sit at 92.6\% and 93.3\% at $T{=}0$, so sampling does not raise their rates.}
\label{tab:temp}
\end{table*}

\section{Five-option tasks}
\label{app:fiveopt}

Table~\ref{tab:fiveopt} reports the out-of-sample check of \S\ref{sec:robustness}: 300 five-option tasks generated from the same pool (seed 2027), scored in both orderings at temperature 0 by five LLMs chosen to span the engagement spectrum, with the paper's binary-estimated utilities predicting each five-way pick. All five LLMs beat chance by a wide margin on every metric. Position use separates cleanly: the three engaged LLMs place 2--10\% of picks on the first slot, while the boundary LLM Mistral-Nemo 12B and the locked Llama 3.1 8B keep following it even with five options on screen.

\begin{table*}[!t]
\centering
\small
\begin{tabular}{lrrrrr}
\toprule
LLM & First-slot & Hit & Top-2 & Mean pred.\ rank & $\beta_{\ln p}$ five-opt (SE) \\
\midrule
Gemma2 9B & 10.2\% & 52.2\% & 84.2\% & 1.67 & $-3.49$ (0.25) \\
Qwen3 4B & 1.8\% & 50.7\% & 78.0\% & 1.80 & $-2.61$ (0.21) \\
Phi-4 Mini & 6.3\% & 44.3\% & 69.8\% & 2.04 & $-0.48$ (0.15) \\
Mistral-Nemo 12B & 55.5\% & 46.7\% & 67.7\% & 2.12 & $-1.82$ (0.18) \\
Llama3.1 8B & 61.2\% & 42.3\% & 66.2\% & 2.12 & $-1.81$ (0.17) \\
\bottomrule
\end{tabular}
\caption{Five-option pilot ($n = 600$ responses per LLM). First-slot = share of picks on the first-listed option; Hit = the binary-estimated utilities' top prediction is chosen (chance 20\%); Top-2 = the choice falls in the predicted top two (chance 40\%); mean predicted rank of the chosen option has chance value 3.0. The last column re-fits the log-price coefficient on the five-option choices themselves.}
\label{tab:fiveopt}
\end{table*}

\section{Bootstrap intervals and multiplicity}
\label{app:bootstrap}

Ratios of noisy coefficients need more than delta-method standard errors, so we re-estimated everything on 500 cluster-bootstrap draws that resample the 450 unique binary choice pairs whole, applying identical draws to every LLM so per-draw rankings are comparable; none of the $500 \times 23 = 11{,}500$ re-fits failed. Table~\ref{tab:bootstrap} reports per-LLM 95\% percentile intervals for the log-price coefficient and the willingness to pay, and the corresponding rank intervals.

Rankings are stable: the mean Spearman correlation between a draw's ranking and the reported one is 0.996 for price sensitivity (23 engaged LLMs) and 0.991 for willingness to pay (20 well-identified LLMs); Phi-4 Mini ranks first on willingness to pay in 99.6\% of draws, Gemma2 9B lands in the bottom two in 100\%, and GPT-5.4 in the bottom three in 99.0\%. The 16$\times$ WTP spread has a 95\% interval of $[11.7\times, 23.7\times]$ (median 15.7$\times$). Of the 190 pairwise willingness-to-pay comparisons among the 20 LLMs, 148 (78\%) have non-overlapping intervals: adjacent LLMs are often not separable, the extremes and the spread are.

For multiplicity, we audited all 184 coefficient tests (8 coefficients $\times$ 23 engaged LLMs) under Bonferroni and Benjamini--Hochberg corrections at $\alpha = 0.05$: 155 are significant at raw $p < 0.05$, 138 survive Bonferroni, 155 survive Benjamini--Hochberg. Among the 40 coefficients the paper claims at $p < 0.001$ (price and review score for the 20-LLM screen), the largest raw $p$ is $7.73\times10^{-7}$ and none loses significance under either correction.

\begin{table*}[!t]
\centering
\small
\begin{tabular}{lrrcc}
\toprule
LLM & $\beta_{\ln p}$ [95\% CI] & WTP [95\% CI] & Rank $|\beta_{\ln p}|$ [CI] & Rank WTP [CI] \\
\midrule
GPT-5.4 & $-7.02$ [$-8.08$, $-6.29$] & 107 [94, 121] & 1 [1, 1] & 19 [18, 19] \\
Gemma2 9B & $-5.71$ [$-6.24$, $-5.27$] & 68 [59, 78] & 2 [2, 3] & 20 [20, 20] \\
GPT-4.1 Mini & $-5.68$ [$-6.35$, $-5.15$] & 119 [108, 131] & 3 [2, 3] & 18 [17, 19] \\
Claude Haiku 4.5 & $-4.75$ [$-5.28$, $-4.31$] & 139 [122, 157] & 4 [4, 6] & 16 [16, 17] \\
GPT-5.4 Mini & $-4.56$ [$-5.18$, $-4.11$] & 222 [203, 246] & 5 [4, 6] & 12 [11, 14] \\
Gemma3 27B & $-4.38$ [$-4.91$, $-3.92$] & 210 [191, 233] & 6 [4, 6] & 13 [12, 15] \\
Phi-4 14B & $-3.27$ [$-3.69$, $-2.94$] & 202 [181, 228] & 7 [7, 7] & 15 [13, 15] \\
GPT-5.4 Nano & $-2.47$ [$-2.86$, $-2.13$] & 500 [446, 572] & 8 [8, 10] & 5 [3, 5] \\
GPT-4.1 Nano & $-2.42$ [$-2.77$, $-2.12$] & 417 [377, 469] & 9 [8, 10] & 6 [5, 8] \\
Phi-3 Medium & $-2.18$ [$-2.48$, $-1.94$] & 276 [240, 309] & 10 [9, 11] & 10 [10, 11] \\
Gemma3 12B & $-2.15$ [$-2.45$, $-1.92$] & 410 [367, 461] & 11 [9, 11] & 7 [5, 8] \\
Qwen3 4B & $-1.59$ [$-1.80$, $-1.41$] & 127 [109, 147] & 12 [12, 13] & 17 [16, 19] \\
Mistral-Nemo 12B & $-1.37$ [$-1.54$, $-1.24$] & 203 [179, 233] & 13 [13, 14] & 14 [12, 15] \\
Gemma3 4B & $-1.33$ [$-1.55$, $-1.11$] & 337 [291, 396] & 14 [13, 14] & 9 [8, 10] \\
DeepSeek-R1 1.5B & $-0.96$ [$-1.12$, $-0.83$] & --- & 15 [15, 17] & --- \\
Llama3.2 3B & $-0.92$ [$-1.14$, $-0.71$] & 525 [428, 656] & 16 [15, 17] & 3 [2, 6] \\
Phi-4 Mini & $-0.76$ [$-1.04$, $-0.50$] & 1071 [819, 1574] & 17 [15, 19] & 1 [1, 1] \\
DeepSeek-R1 7B & $-0.67$ [$-0.83$, $-0.51$] & 340 [284, 438] & 18 [17, 19] & 8 [6, 9] \\
Qwen3 8B & $-0.55$ [$-0.67$, $-0.44$] & 263 [220, 321] & 19 [18, 20] & 11 [10, 12] \\
Phi-3 Mini & $-0.47$ [$-0.61$, $-0.32$] & 646 [501, 926] & 20 [19, 21] & 2 [2, 4] \\
Qwen3 30B-A3B & $-0.40$ [$-0.50$, $-0.30$] & 511 [414, 677] & 21 [20, 21] & 4 [2, 6] \\
Gemma3 1B & $+0.13$ [$-0.04$, $+0.28$] & --- & 22 [22, 23] & --- \\
Phi-2 2.7B & $-0.07$ [$-0.22$, $+0.09$] & --- & 23 [22, 23] & --- \\
\bottomrule
\end{tabular}
\caption{Cluster-bootstrap 95\% percentile intervals ($B = 500$, resampling the 450 binary choice pairs whole) for the log-price coefficient and the willingness to pay for one review-score point at the \$250 reference (dollars), with rank intervals. Price-sensitivity ranks cover the 23 engaged LLMs (1 = most price-sensitive); WTP ranks cover the 20-LLM well-identified screen (1 = highest WTP). The three LLMs outside the screen (not both coefficients significant at $p < 0.001$) have no meaningful WTP and are dashed out.}
\label{tab:bootstrap}
\end{table*}

\section{Human-anchor conversion arithmetic}
\label{app:anchor}

This appendix documents every conversion behind the human reference range of \S\ref{sec:results_human}. The LLM benchmark values are the 20 well-identified engaged LLMs' willingness to pay for one point on the 10-point review scale at the \$250 nightly reference: \$68 (Gemma2 9B) to \$1{,}071 (Phi-4 Mini), or 27.4\% to 428.6\% of the nightly rate, with a median of \$269 (107.7\%).

Three scale conversions recur. A point on an OTA's 5-point review scale spans the same range as two points on the 10-point scale, so per-point effects on a 5-point scale are halved. ReviewPro's Global Review Index (GRI) is a 0--100 reputation score whose effect is reported per 1\% increase; multiplying by ten approximates one 10-point-scale point, a linear extrapolation that makes the result an upper bound. Hong Kong dollars convert at the 7.80 HKD/USD peg. Table~\ref{tab:anchor} applies these to each source; all dollar conversions are ours, computed from the published coefficients.

\begin{table*}[!t]
\centering
\small
\setlength{\tabcolsep}{3pt}
\begin{tabular}{p{3.8cm}p{4.4cm}p{4.2cm}p{2.5cm}}
\toprule
Study & Setting and construct & Conversion & Per point at \$250 \\
\midrule
\citet{anderson2012lodging} (research-center report) & Travelocity purchase logit; price change holding purchase probability constant & 11.2\% per 5-pt point $\div$ 2 = 5.6\% & \$14.00 \\
\citet{anderson2012lodging}, same report & GRI-to-ADR equilibrium regression & 0.89\% per 1\% of GRI $\times$ 10 = 8.9\% (upper bound) & up to \$22.25 \\
\citet{lewis2016welfare} (working paper) & Five-state hotel panel; price regression, hotel fixed effects & 1.5\% (pooled) to 9\% (2014) per 5-pt point $\div$ 2 & \$1.88--\$11.25 \\
\citet{ursu2018rankings} & Expedia, randomized ranking; sequential search model & review coefficient negative or insignificant, all four destinations & $\leq$ \$0 \\
\citet{ghose2012ranking} & Travelocity transactions; random-coefficients logit, log price & ratio of review-rating to log-price coefficients (our derivation; the conversion equation of \citealp{ghose2014examining} independently corroborates the ratio, 0.237 against 0.228) & \$29--\$33 per platform; \$61 summed \\
\citet{chen2017sequential} & US OTA clickstream; dollar-denominated utility & entire above-4.5 versus below-4 band worth \$99.11 per night & $\leq$ \$99 (band, not point) \\
\citet{masiero2015wtp} & Stated-choice experiment, Hong Kong guests & largest single-attribute WTP HK\$771 (harbour view) $\div$ 7.80 & \$99 (attribute, not point) \\
\citet{pawlicz2017determinants} & Warsaw hedonic price model & 25--36\% per official star & \$62.50--\$90.00 per star \\
\bottomrule
\end{tabular}
\caption{Human anchors for the value of one review-score point (10-point scale) at a \$250 nightly rate. The first three rows are equilibrium price movements; the next three are structural estimates from real bookings; the last two are stated-choice and hedonic context at coarser units. Demand-side evidence \citep{ogut2012reviews,viglia2016ewom} supports the same direction but is not convertible to dollars per point.}
\label{tab:anchor}
\end{table*}

Two features of the human evidence matter for reading the comparison. First, the human benchmark is a distribution, not a number: household price coefficients in scanner panels vary by nearly half their average level with demographics explaining only 7\% of the variance \citep{rossi1996value}, taste dispersion rivals or exceeds mean tastes in \citet{berry1995automobile}, and tourist segments differ in price sensitivity by roughly $4\times$, including a 28\% segment whose price coefficient is essentially zero (our computation from Table~3 of \citealp{masiero2012segmentation}). The LLM population reproduces this structure at larger magnitude, scale and taste varying separately \citep{fiebig2010gmnl}. Second, the constructs differ from ours: equilibrium price premia reflect supply as well as demand and are attenuated toward the marginal consumer; stated choices carry hypothetical-inflation caveats; the search-model estimates condition on rich controls in environments with search costs, whereas PriceBench shows fully described options side by side. Humans also spend their own money under budget constraints, while LLM agents book with synthetic money, a plausible mechanism for inflated quality valuations and precisely why agents should be audited before delegation. The anchors bound plausibility; they do not define a correct value.

\section{Datasheet}
\label{app:datasheet}

We document PriceBench following the \emph{Datasheets for Datasets} framework \citep{gebru2021datasheets}.

\paragraph{Motivation.} PriceBench was created to measure the price, quality, and brand preferences an LLM reveals when it chooses among qualifying options as a booking agent, rather than whether it completes the task.

\paragraph{Composition.} The benchmark has two parts. (i) A frozen pool of 179 real New York City hotel profiles spanning 1--5 stars, 36 neighborhoods, five hotel chains plus independents, and \$45--\$1{,}650 per night; each profile carries ten attributes (four fixed: star rating, neighborhood, chain, amenities; six re-randomized per appearance: price, room type, free-cancellation flag, breakfast flag, guest review score, review count). (ii) 3{,}600 forced-choice tasks built from the pool: 450 unique pairs each shown 4 times (1{,}800 binary) and 300 unique triples each shown 6 times (1{,}800 ternary). The release also includes the response sets: 28 LLMs from 8 providers, each task presented in both orderings, up to 7{,}200 observations per LLM. Instances are commercial hotel listings; no personal or human-subject data is included.

\paragraph{Collection.} Hotel profiles were compiled from public listings on four online travel agencies (Booking.com, Expedia, KAYAK, TripAdvisor) and then frozen; the data describes lodging, not individuals. Model responses were collected by prompting each LLM with plain-text option cards (Appendix~\ref{app:prompt}) at temperature 0 with greedy decoding.

\paragraph{Preprocessing and cleaning.} On each appearance, six attributes are re-randomized so repeated presentations are non-identical: price drawn uniformly within the property's listed range, review score perturbed within $\pm 0.2$, review count within $\pm 10\%$, and room type, cancellation, and breakfast flags resampled. Perturbations are confined to each property's listed range so every profile stays a plausible real listing. Task generation uses a fixed seed (2026); price deciles for the non-parametric analysis are computed once on the pooled price distribution.

\paragraph{Uses.} PriceBench is intended for measuring and comparing the revealed preferences of LLM booking agents and for discrete-choice analysis of model behavior. It is a diagnostic instrument, not a source of booking recommendations, and the perturbed prices and reviews should not be treated as current commercial information.

\paragraph{Distribution.} The tasks, scoring code, analysis pipeline, and all response sets are released publicly at \url{https://github.com/Pashasan/pricebench-emnlp}, together with the prompt-variant, decoding-temperature, and five-option runners of \S\ref{sec:robustness} and the 300 five-option tasks. Code is under the MIT License; the task and response data are released for research use. The hotel pool is a derived research artifact built from public listings with re-randomized prices and review fields; users should consult the source platforms' terms before any non-research reuse.

\paragraph{Maintenance.} The benchmark is versioned with the code release and maintained by the author; a new LLM is added by scoring it twice and registering the response pair, after which the full analysis reproduces automatically.

\end{document}